\documentclass[twocolumn,showpacs,superscriptaddress,prb,amsmath,amssymb,floatfix,eqsecnum]{revtex4}
\usepackage{amsmath,amssymb}
\usepackage{bm}
\usepackage{epsfig,psfrag}

\newcommand{\G}{\mathcal{G}}
\newcommand{\Z}{\mathcal{Z}}
\newcommand{\LL}{\mathcal{L}}
\newcommand{\idf}{\int\!\!D\Phi\,\,}

\newcommand{\Go}{\mathbf{G}_0}
\newcommand{\p}[1]{\frac{\delta}{\delta #1}}
\newcommand{\pp}[2]{\frac{\delta #1}{\delta #2}}
\newcommand{\ppt}[3]{\frac{\delta^{(2)} #1}{\delta #2 \delta #3}}
\newcommand{\Tr}{\mathrm{Tr}}

\begin{document}

\title{Collective fields in the functional renormalization group for
  fermions, Ward identities, and the exact solution of the
  Tomonaga-Luttinger model}
  
\author{Florian Sch\"{u}tz} 
\affiliation{Institut f\"{u}r Theoretische
  Physik, Universit\"{a}t Frankfurt, Max-von-Laue-Strasse 1, 60054
  Frankfurt, Germany}

\author{Lorenz Bartosch} 
\affiliation{Institut f\"{u}r Theoretische
  Physik, Universit\"{a}t Frankfurt, Max-von-Laue-Strasse 1, 60054
  Frankfurt, Germany} 

\affiliation{Department of Physics, Yale University, P.O.Box 208120,
  New Haven, CT 06520-8120, USA}
  
\author{Peter Kopietz} 
\affiliation{Institut f\"{u}r Theoretische
  Physik, Universit\"{a}t Frankfurt, Max-von-Laue-Strasse 1, 60054
  Frankfurt, Germany}

\date{September 17, 2004}

\begin{abstract}
  We develop a new formulation of the functional renormalization group
  (RG) for interacting fermions. Our approach unifies the purely
  fermionic formulation based on the Grassmannian functional integral,
  which has been used in recent years by many authors, with the
  traditional Wilsonian RG approach to quantum systems pioneered by
  Hertz [Phys. Rev. B {\bf 14}, 1165 (1976)], which attempts to
  describe the infrared behavior of the system in terms of an
  effective bosonic theory associated with the soft modes of the
  underlying fermionic problem.  In our approach, we decouple the
  interaction by means of a suitable Hubbard-Stratonovich
  transformation (following the Hertz-approach), but do not eliminate
  the fermions; instead, we derive an exact hierarchy of RG flow
  equations for the irreducible vertices of the resulting coupled
  field theory involving both fermionic and bosonic fields.  The
  freedom of choosing a momentum transfer cutoff for the bosonic soft
  modes in addition to the usual band cutoff for the fermions opens
  the possibility of new RG schemes.  In particular, we show how the
  exact solution of the Tomonaga-Luttinger model (i.e.,
  one-dimensional fermions with linear energy dispersion and
  interactions involving only small momentum transfers) emerges from
  the functional RG if one works with a momentum transfer cutoff.
  Then the Ward identities associated with the local particle
  conservation at each Fermi point are valid at every stage of the RG
  flow and provide a solution of an infinite hierarchy of flow
  equations for the irreducible vertices.  The RG flow equation for
  the irreducible single-particle self-energy can then be closed and
  can be reduced to a linear integro-differential equation, the
  solution of which yields the result familiar from bosonization.  We
  suggest new truncation schemes of the exact hierarchy of flow
  equations, which might be useful even outside the weak coupling
  regime.
\end{abstract}


\maketitle

\section{Introduction}

In condensed matter and statistical physics the renormalization group
(RG) in the form developed by Wilson and coauthors
\cite{Wilson72,Wilson74,Fisher98} has been very successful.  At the
heart of this intuitively appealing formulation of the RG lies the
concept of an effective action describing the physical properties of a
system at a coarse grained scale.  Conceptually, the derivation of
this effective action is rather simple provided the theory can be
formulated in terms of a functional integral: one simply integrates
out the degrees of freedom describing short-wavelength fluctuations in
a certain regime, and subsequently rescales the remaining degrees of
freedom in order to compare the new effective action with the original
one.  Of course, in practice, the necessary functional integration can
almost never be performed analytically, so that one has to resort to
some approximate procedure.\cite{Ma76} However, if the elimination of
the short wavelength modes is performed in infinitesimal steps, one
can write down a formally exact RG flow equation describing the change
of the effective action due to mode elimination and rescaling. The
earliest version of such a functional RG equation has been derived by
Wegner and Houghton.\cite{Wegner73} Subsequently, many authors have
derived alternative versions of the functional RG with the same
physical content, using different types of generating functionals.  In
particular, the advantages of working with the generating functional
of the one-particle irreducible vertices have been realized early on
by Di Castro, Jona-Lasinio, and Peliti,\cite{DiCastro74} and by
Nicoll, Chang, and Stanley.\cite{Nicoll76,Chang92} The focus of the
above works was an accurate description of second-order phase
transitions at finite temperatures, where quantum mechanics is
irrelevant.

In recent years, there has been much interest in quantum systems
exhibiting phase transitions as a function of some nonthermal control
parameter, such as pressure or density.  In a pioneering paper, Hertz
\cite{Hertz76} showed how the powerful machinery of the Wilsonian RG
can be generalized to study quantum critical phenomena in Fermi
systems.  Technically, this is achieved with the help of so-called
Hubbard-Stratonovich transformations, which replace the fermionic
two-particle interaction by a suitable bosonic field that couples to a
quadratic form in the fermion operators.\cite{Negele88} The fermions
can then be integrated out in a formally exact way, resulting in an
effective action for the bosonic field.  Of course, there are many
possible ways of decoupling fermionic two-body interactions by means
of Hubbard-Stratonovich transformations.  In the spirit of the usual
Ginzburg-Landau-Wilson approach to classical critical phenomena, one
tries to construct the effective bosonic theory such that the field
can be identified with the fluctuating order parameter, or its field
conjugate.  However, if the system supports additional soft modes that
couple to the order parameter,\cite{Kirkpatrick96,Rosch01} an attempt
to construct an effective field theory in terms of the order parameter
alone leads in general to an effective action with singular and
nonlocal vertices, so that the usual RG methods developed for
classical phase transitions cannot be applied.  In this case, it is
better to construct an effective action involving all soft modes
explicitly.  However, for a given problem the nature of the soft modes
is not known {\it a priori}, so that the explicit introduction of the
corresponding degrees of freedom by means of a suitable
Hubbard-Stratonovich transformation is always based on some prejudice
about the nature of the ground state and the low-lying excitations of
the system.  Recently, the breakdown of simple Ginzburg-Landau-Wilson
theory has also been discussed in the context of quantum
antiferromagnets by Senthil {\em et~al.}\cite{Senthil04}

In order to have a completely unbiased RG approach for interacting
fermions, one can also apply the Wilsonian RG directly to the
Grassmannian functional integral representation of the partition
function or the Green's functions of an interacting Fermi
system.\cite{Shankar94} In the past ten years, several groups have
further developed this
method.\cite{Zanchi96,Halboth00,Honerkamp01,Honerkamp01b,Salmhofer01,Kopietz01,Busche02,Ledowski03,Tsai01,Binz02,Meden02,Kampf03,Katanin04}
In particular, the mode elimination step of the RG transformation has
been elegantly cast into formally exact differential equations for
suitably defined generating functionals.  These functional equations
then translate to an infinite hierarchy of integro\-differential
equations for the vertices.  On a technical level, it is often
advantageous to work with the generating functional of the
one-particle irreducible vertices.\cite{DiCastro74,Nicoll76,Chang92}
For classical field theories the exact RG flow equation for this
generating functional has been obtained by
Wetterich,\cite{Wetterich93} and by Morris.\cite{Morris94} For
nonrelativistic fermions, the corresponding flow equation has been
derived by Salmhofer and Honerkamp \cite{Salmhofer01} and,
independently and in more explicit form, by Kopietz and
Busche.\cite{Kopietz01}

However, the purely fermionic formulation of the Wilsonian RG has
several disadvantages. In order to obtain at least an approximate
solution of the formally exact hierarchy of RG flow equations for the
vertices, severe truncations have to be made, which can only be
justified as long as the fermionic four-point vertex (i.e., the
effective interaction) remains small.  Hence, in practice the
fermionic functional RG is restricted to the weak coupling regime, so
that possible strong coupling fixed points are not accessible within
this method.  Usually, one-loop truncated RG equations are iterated
until at least one of the marginal interaction constants becomes
large, which is then interpreted as a weak-coupling instability of the
Fermi system in the corresponding
channel.\cite{Halboth00,Honerkamp01,Honerkamp01b}

Within the framework of the Wilsonian functional RG, a consistent
two-loop calculation has not been performed so far due to the immense
technical difficulties involved.  Such a calculation should also take
into account the frequency dependence of the effective interaction and
the self-energy corrections to the internal Green's functions.  Note
that in the work by Katanin and Kampf \cite{Kampf03} the frequency
dependence of the effective interaction has been ignored, so that the
effect of possible bosonic collective modes is not included in these
calculations.  It is well known that two-loop calculations are more
conveniently performed using the field-theoretical RG method, provided
the physical problem of interest can be mapped onto a renormalizable
field theory.  Ferraz and coauthors \cite{Ferraz03,Ferraz03b}
recently used the field-theoretical RG to calculate the
single-particle Green function at the two loop level for a special
two-dimensional Fermi system with a flat Fermi surface. Interestingly,
they found a new non-Fermi liquid fixed point characterized by a
finite renormalized effective interaction and a vanishing
wave-function renormalization.  The running interaction without
wave-function renormalization diverges in this model, but the
divergence is canceled by the vanishing wave-function renormalization
such that the renormalized interaction remains finite.

Another difficulty inherent in any RG approach to Fermi systems at
finite densities arises from the fact that the true Fermi surface of
the interacting many-body system is not known {\it a priori}. In fact, in
dimensions $D > 1$ interactions can even change the symmetry of the
Fermi surface.\cite{Pomeranchuk58,Metzner04} In a perturbative
approach, finding the renormalized Fermi surface is a delicate
self-consistency problem \cite{Nozieres64}; if one starts from the
wrong Fermi surface one usually encounters unphysical
singularities.\cite{Kohn60} So far, the self-consistent
renormalization of the Fermi surface has not been included in the
numerical analysis of the one-loop truncated fermionic functional RG.
As shown in Refs.~\onlinecite{Kopietz01}, \onlinecite{Ledowski03}, and
\onlinecite{Ferraz03b}, the renormalized Fermi surface can be defined
as a fixed point of the RG and can in principle be calculated
self-consistently entirely within the RG framework.

The progress in overcoming the above difficulties inherent in the RG
approach to fermions using a purely fermionic parametrization has
been rather slow.  In our opinion, this has a simple physical reason:
the low lying excitations (i.e., soft modes) of an interacting Fermi
system consist not only of fermionic quasiparticles, but also of
bosonic collective excitations.\cite{Pines89} The latter are rather
difficult to describe within the purely fermionic parametrizations
used in
Refs.~\onlinecite{Zanchi96,Halboth00,Honerkamp01,Honerkamp01b,Salmhofer01,Kopietz01,Busche02,Ledowski03,Tsai01,Binz02,Meden02,Kampf03,Katanin04}.
Naturally, a formulation of the functional RG where both fermionic and
bosonic excitations are treated on equal footing should lead to a more
convenient parametrization.  Such a strategy seems also natural in
light of the observation by Kirkpatrick, Belitz, and co-workers
\cite{Kirkpatrick96} (see also Ref.~\onlinecite{Rosch01}) that an
effective low-energy and long-wavelength action with well-behaved
vertices is only obtained if the soft modes are not integrated out,
but appear explicitly as quantum fields.

Our aim in this work is to set up a functional renormalization group
scheme that allows for a simultaneous treatment of fermionic as well
as collective bosonic degrees of freedom. This is done by explicitly
decoupling the interaction via a Hubbard-Stratonovich transformation
in the spirit of Hertz \cite{Hertz76} and then considering the
functional renormalization group equations for the mixed field theory
involving both fermionic and bosonic fields.  This type of approach
has been suggested previously by Correia, Polonyi, and
Richert,\cite{Correia01} who studied the homogeneous electron gas by
means of a gradient expansion of a functional version of a
Callan-Symanzik equation.  Here, we follow the more standard approach
and derive a hierarchy of flow equations for the vertex functions of
our coupled Fermi-Bose theory. A related approach has also been
developed by Wetterich and coauthors.\cite{Wetterich04,Baier03}
However, they discussed only a simple truncation of the exact
hierarchy of RG flow equations involving an effective potential in a
bosonic sector and a momentum-independent Yukawa coupling. They did
not pay any attention to the problem of the compatibility of the RG
flow with Ward identities, which play a crucial role for interacting
fermions with dominant forward scattering.  Our procedure bridges the
gap between the purely bosonic approach to quantum critical phenomena
by Hertz \cite{Hertz76} and the purely fermionic functional RG method
developed in the past decade by several
authors.\cite{Zanchi96,Halboth00,Honerkamp01,Honerkamp01b,Salmhofer01,Kopietz01,Busche02,Ledowski03,Tsai01,Binz02,Meden02,Kampf03,Katanin04}
Contrary to the approach by Hertz,\cite{Hertz76} in our approach the
fermionic degrees of freedom are still present, so that the feedback
of the collective bosonic modes on the one-particle spectral
properties of the fermions can be studied.

The parametrization of the low-lying excitations of an interacting
Fermi system in terms of collective bosonic fields is most natural in
one spatial dimension,\cite{Giamarchi04,Haldane81} where Fermi-liquid
theory breaks down and is replaced by the Luttinger-liquid
concept.\cite{Haldane81} An exactly solvable paradigm of a Luttinger
liquid is the Tomonaga-Luttinger model
(TLM),\cite{Giamarchi04,Haldane81,Stone94} consisting of fermions with
exactly linear energy dispersion and interactions involving only small
momentum transfers. A description in terms of bosonic variables is
well known to provide an exact solution for thermodynamic quantities
as well as correlation functions.\cite{Giamarchi04,Haldane81,Stone94}
One might then wonder if it is also possible to obtain the complete
form of the correlation functions of the TLM entirely within the
framework of the functional RG.  An attempt \cite{Busche02} to
calculate the momentum- and frequency-dependent single-particle
spectral function $A ( k , \omega )$ of the TLM by means of an
approximate iterative two-loop solution of the functional RG equations
at weak coupling yields the correct behavior of $A ( \pm k_F ,
\omega)$ known from bosonization (where $k_F$ is the Fermi momentum),
but yields incorrect threshold singularities for momenta away from
$\pm k_F$.  In this work we go considerably beyond Ref.
\onlinecite{Busche02} and show how the TLM can be solved
{\it{exactly}} using our mixed fermionic-bosonic functional RG.  A
crucial point of our method is that the RG can be set up in such a way
that the Ward identities underlying the exact solubility of the TLM
\cite{Dzyaloshinskii74,Bohr81,Metzner98,Kopietz97} are preserved by
the RG. This is not the case in the purely fermionic formulation of
the functional RG.\cite{Katanin04}

Very recently, Benfatto and Mastropietro\cite{Benfatto04} also
developed an implementation of the RG for the TLM which takes the
asymptotic Ward identities into account. However, these authors did
not introduce bosonic collective fields and did not attempt to
calculate the exact single-particle Green's function.
  
The rest of the paper is organized as follows. In
Sec.~\ref{sec:preliminaries}, we introduce the model, carry out the
decoupling of the interaction and set up a compact notation which
turns out to facilitate the bookkeeping in the derivation of the
functional RG equations that is presented in detail in
Sec.~\ref{sec:flow_eq}. In Sec.~\ref{sec:1dflow}, we
introduce a new RG scheme which uses the momentum transfer of the
effective interaction as a cutoff. We show that for linearized energy
dispersion the resulting infinite hierarchy of flow equations for the
irreducible vertices involving two external fermion legs and an
arbitrary number of external boson legs can be solved exactly by means
of an infinite set of Ward identities.  Using these identities, the
flow equation for the irreducible self-energy can then be reduced to a
closed linear integro\-differential equation, which can be solved
exactly.  In one dimension, we recover in Sec.~\ref{subsec:exactTLM}
the exact solution of the Tomonaga-Luttinger model in the form
familiar from functional bosonization \cite{Kopietz97,Kopietz95}.
Finally, in Sec.~\ref{sec:summary}, we summarize our results and give
a brief outlook on possible further applications of our method.  There
are three appendices where we present some more technical details.  In
Appendix A, we use our compact notation introduced in
Sec.~\ref{sec:preliminaries} to discuss the structure of the tree
expansion in our coupled Fermi-Bose theory. Appendix B contains a
derivation of the skeleton diagrams for the first few irreducible
vertices of our theory using the Dyson-Schwinger equations of motion,
which follow from the invariance of the functional integral with
respect to infinitesimal shift transformations. Finally, in Appendix C
we use the gauge invariance of the mixed Fermi-Bose action to derive a
cascade of infinitely many Ward identities involving vertices with two
fermion legs and an arbitrary number of boson legs.

\section{Interacting Fermions as coupled Fermi-Bose systems}
\label{sec:preliminaries}

In this section we discuss the Hubbard-Stratonovich transformation and
set up a condensed notation to treat fermionic and bosonic fields on
the same footing. This will allow us to keep track of the rather
complicated diagrammatic structure of the flow equations associated
with our coupled Fermi-Bose system in a very efficient way. A similar
notation has been used previously in Refs.~\onlinecite{Salmhofer01}
and \onlinecite{Baier03}.

\subsection{Hubbard-Stratonovich transformation}

We consider a normal fermionic many-body system with two-particle
density-density interactions.  In the usual Grassmannian functional
integral approach \cite{Negele88} the grand-canonical partition
function and all (imaginary)-time-ordered Green's functions can be
represented as functional averages involving the following Euclidean
action:
\begin{eqnarray}
  S[\bar{\psi} , \psi]&=& S_0 [\bar{\psi}, \psi] + S_{\mathrm{int}} [\bar{\psi} ,\psi]
 \label{eq:Spsidef}
 \; ,
 \\ S_0 [\bar{\psi}, \psi]
 && = \sum_{\sigma}\int_{K} \bar{\psi}_{K\sigma}
  [-i\omega + \xi_{{\bf k}\sigma}]\psi_{K\sigma}
  \label{eq:S0psidef}
 \; ,
 \\
 S_{\mathrm{int}} [\bar{\psi} ,\psi]
  & = &
  \frac12\sum_{\sigma\sigma'}\int_{\bar{K}}
  f_{\bar{\bf{k}}}^{\sigma\sigma'}
  \bar{\rho}_{\bar{K}\sigma}\rho_{\bar{K}\sigma'}\,,
 \label{eq:Sintpsidef}
\end{eqnarray}
where the composite index $K=(i\omega,{\bf k})$ contains a fermionic
Matsubara frequency $i \omega$ as well as an ordinary wave vector
${\bf{k}}$. Here the energy dispersion
$\xi_{\mathbf{k}\sigma}=\epsilon_{\mathbf{k}\sigma}-\mu$ is measured
relative to the chemical potential $\mu$, and
$f_{\bar{\bf{k}}}^{\sigma\sigma'}$ are some momentum-dependent
interaction parameters.  Throughout this paper, labels with an overbar
refer to bosonic frequencies and momenta, while labels without an overbar
refer to fermionic ones.  We have normalized the Grassmann fields
$\psi_{K \sigma}$ and $\bar{\psi}_{K \sigma}$ such that the
integration measure in Eq.~(\ref{eq:Spsidef}) is
\begin{equation}
  \int_K = \frac{1}{\beta V}\sum_{\omega, {\bf{k}}} 
  \quad\stackrel{\beta,V\to\infty}{\longrightarrow}\quad
  \int\frac{d\omega}{2\pi}\frac{d^Dk}{(2\pi)^D}\,,
\end{equation}
where $\beta$ is the inverse temperature and $V$ is the volume of the
system. The Fourier components of the density are represented by the
following composite field:
\begin{equation}
  \rho_{\bar{K}\sigma}=\int_K
  \bar{\psi}_{K \sigma}\psi_{K+ \bar{K},\sigma}\,,
\end{equation}
which implies $\bar{\rho}_{\bar{K}\sigma}=\rho_{-\bar{K}\sigma}$. The
discrete index $\sigma$ is formally written as a spin projection, but
will later on also serve to distinguish right and left moving fields
in the Tomonaga-Luttinger model. This is why a dependence of the
dispersion $\xi_{{\bf{k}} \sigma} $ on $\sigma$ has been kept.

The interaction is bilinear in the densities and can be decoupled by
means of a Hubbard-Stratonovich transformation.\cite{Kopietz97} The
interaction is then mediated by a real field $\varphi$ and the
resulting action reads as
\begin{eqnarray}
  S[\bar{\psi}, \psi,\varphi]&=& S_0 [\bar{\psi}, \psi] + S_0 [\varphi]+
  S_{1}[\bar{\psi}, \psi,\varphi]
  \; ,
  \label{eq:Spsiphidef}
\end{eqnarray}
where the free bosonic part is given by
\begin{eqnarray}
  S_0 [\varphi] &=&
  \frac12\sum_{\sigma\sigma'}\int_{\bar{K}} 
  [f^{-1}_{\bar{\bf{k}}}]^{\sigma\sigma'}
  \varphi^*_{\bar{K}\sigma} \varphi_{\bar{K}\sigma'} 
  \label{eq:S0phi}
  \; ,
\end{eqnarray}
and the coupling between Fermi and Bose fields is
\begin{eqnarray}
  S_{1}[\bar{\psi} , \psi,\varphi] &=&
  i\sum_{\sigma}\int_{\bar{K}}
  \bar{\rho}_{\bar{K}\sigma}\varphi_{\bar{K}\sigma}
  \nonumber \\
  & = &
  i\sum_{\sigma}\int_{{K}} \int_{\bar{K}}
  \bar{\psi}_{ K + \bar{K}, \sigma} \psi_{K \sigma} \varphi_{\bar{K}\sigma}
  \,.
  \label{eq:Spsiphi}
\end{eqnarray}
The Fourier components of a real field satisfy
$\varphi^*_{\bar{K}\sigma}=\varphi_{-\bar{K}\sigma}$.  For the
manipulations in the next section it will prove advantageous to
further condense the notation and collect the fields in a vector $\Phi
= (\psi,\bar{\psi},\varphi)$. The quadratic part of the action can
then be written in the symmetric form
\begin{eqnarray}
  S_0[\Phi]& = & S_0[\bar{\psi}, \psi]+S_0 [\varphi] =
 -\frac12\left(\Phi,\left[{\bf G}_0\right]^{-1}\Phi\right)
 \nonumber
 \\
  & = &
  - \frac12 \int_{\alpha} \int_{\alpha^{\prime}} \Phi_{\alpha}
\left[{\bf G}_0\right]^{-1}_{\alpha \alpha^{\prime}} \Phi_{ \alpha^{\prime}}
\,,
\end{eqnarray}
where ${\bf G}_0$ is now a matrix in frequency, momentum, spin, and
field-type indices, and $\alpha$ is a ``super label'' for all of these
indices.  The symbol $\int_{\alpha}$ denotes integration over the
continuous components and summation over the discrete components of
$\alpha$.  The matrix ${\bf{G}}_0^{-1}$ has the block structure
\begin{equation}
  \mathbf{G}_0^{-1} = \left(\begin{array}{ccc}
    0&\zeta [\hat{G}_0^{-1}]^T&0\\
    \hat{G}_0^{-1}&0&0\\
    0&0&-\hat{F}_0^{-1}
  \end{array}\right)\,,
 \label{eq:G0matrixinv}
\end{equation}
where~\cite{footnotezeta} $\zeta = -1$ and $\hat{G}_0$ and $\hat{F}_0$
are infinite matrices in frequency, momentum, and spin space, with
matrix elements
\begin{eqnarray}
  [ \hat{G}_0 ]_{ K\sigma, K'\sigma'}  &=& \delta_{K,K'} \delta_{\sigma \sigma'}
  G _{0, \sigma} ( K  ) \;  ,
  \label{eq:G0matrixdef}\\
  {}[\hat{F}_0 ]_{ \bar{K} \sigma, \bar{K}^{\prime} \sigma^{\prime}}  &=&  
  \delta_{\bar{K} + \bar{K}', 0 }    
  F_{0, \sigma  \sigma^{\prime} } ( \bar{K} )
  \label{eq:F0matrixdef}
  \; ,
\end{eqnarray}
where
\begin{eqnarray}
  G_{0 , \sigma} ( K ) &=& 
  [i\omega - \xi_{{\bf{k}}\sigma}]^{-1} \,,
  \label{eq:G0def}\\
  F_{0,  \sigma   \sigma^{\prime}} (\bar{K})     
  &=& 
  f_{\bar{\bf{k}}}^{\sigma \sigma'}  \, .
  \label{eq:F0def}
\end{eqnarray}
The Kronecker $\delta_{K,K'}=\beta V
\delta_{\omega,\omega'}\delta_{\mathbf{k},\mathbf{k}'}$ appearing in
Eqs.~(\ref{eq:G0matrixdef},\ref{eq:F0matrixdef}) is normalized such
that it reduces to Dirac $\delta$ functions
$\delta_{K,K'}\to(2\pi)^{D+1}\delta(\omega-\omega')\delta^{(D)}(\mathbf{k}-\mathbf{k}')$
in the limit $\beta,V\to\infty$.  Note that the bare interaction plays
the role of a free bosonic Green's function.  For later reference, we
note that the inverse of Eq.~(\ref{eq:G0matrixinv}) is
\begin{equation}
  \mathbf{G}_0 = \left(\begin{array}{ccc}
    0& \hat{G}_0  &0\\
    \zeta \hat{G}_0^{T}&0&0\\
    0&0&-\hat{F}_0
  \end{array}\right)\,,
 \label{eq:G0matrix}
\end{equation}
and that the transpose of $\mathbf{G}_0$ satisfies
\begin{equation}
  \mathbf{G}_0^T = \mathbf{Z} \mathbf{G}_0  = \mathbf{G}_0 \mathbf{Z}
  \; ,
  \label{eq:G0transpose}
\end{equation}
where the ``statistics matrix'' $\mathbf{Z}$ is defined by
\begin{equation}
  [ \mathbf{Z} ]_{\alpha \alpha^{\prime} } = \delta_{\alpha \alpha^{\prime} }
  \zeta_{\alpha}
  \; .
\end{equation} 
Here, $\zeta_{\alpha} = -1$ if the superindex $\alpha$ refers to a
Fermi field, and $\zeta_{\alpha} = 1$ if $\alpha$ labels a Bose field.

\subsection{Generating functionals}

\subsubsection{Generating functional of connected Green's functions}

We now introduce sources $J_{\alpha}$ and define the generating
functional $\G [J]$ of the Green's functions as follows:
\begin{equation}
  \G[J] = e^{\G_c[J]} = \frac{1}{\Z_0}\idf e^{-S_0-S_1+(J,\Phi)}
 \; .
 \label{eq:Ggen}
\end{equation}
Here, $\G_c [J]$ is the generating functional for connected Green
functions and the partition function $\Z_0$ of the noninteracting
system can be written as the Gaussian integral
\begin{equation}
  \Z_0 = \idf e^{-S_0}\,.
\end{equation}
Let us use the compact notation
\begin{equation}
  (J,\Phi ) = \int_{\alpha} J_{\alpha} \Phi_{\alpha}
  \; .
  \label{eq:JPhiscalarproduct}
\end{equation}
Conventionally, the source terms for fields of different types are
written out explicitly in the form\cite{Negele88}
\begin{eqnarray}
  &&(J,\Phi)=(\bar{\jmath},\psi)+(\bar{\psi},j)+(J^*,\varphi)=
  \nonumber\\
  &&\sum_{\sigma}\int_K \bar{\jmath}_{K\sigma}\psi_{K\sigma}
  + \sum_{\sigma}\int_K \bar{\psi}_{K\sigma}j_{K\sigma}
  + \sum_{\sigma}\int_{\bar{K}} J^*_{\bar{K}\sigma}\varphi_{\bar{K}\sigma}\; .\nonumber\\
  \label{eq:sourcesstandard}
\end{eqnarray}
A comparison between Eq.~(\ref{eq:JPhiscalarproduct}) and
Eq.~(\ref{eq:sourcesstandard}) shows that the sources in the compact
notation are related to the standard ones by $J=(\bar{\jmath},\zeta
j,J^*)$.  The connected $n$-line Green's functions $\G^{(n)}_{c,
  \alpha_1 \ldots \alpha_n} $ are then defined via the functional
Taylor expansion 
\begin{equation}
  \G_c[J]=\sum_{n=0}^{\infty}\frac1{n!}\int_{\alpha_1}\dots
  \int_{\alpha_n} \G^{(n)}_{c, \alpha_1 \dots \alpha_n} J_{\alpha_1}
  \cdot\ldots\cdot J_{\alpha_n}\,,
  \label{eq:Gcexpansion}
\end{equation}
implying
\begin{equation}
  \G^{(n)}_{c, \alpha_1 \dots \alpha_n} =
  \left.
 \frac{ \delta^{(n)} \G_c[J] }{ \delta J_{ \alpha_n} \ldots \delta J_{\alpha_1} }
\right|_{ J = 0 }
\label{eq:Gcndef}
\; .
\end{equation}
In particular, the exact Green's function of our interacting system is
given by
\begin{equation}
  [ \mathbf{G} ]_{\alpha \alpha^{\prime}} =-\left.
    \ppt{\G_c}{J_{\alpha}}{J_{\alpha^{\prime}}}\right|_{J=0} 
  = - \G^{(2)}_{c , \alpha^{\prime} \alpha }
  \,,
  \label{eq:Gmatrix}
\end{equation}
which we shall write in compact matrix notation as
\begin{equation}
  \mathbf{G}=-\left.\ppt{\G_c}{J}{J}\right|_{J=0}
  = 
  \left(\begin{array}{ccc}
      0& \hat{G}  &0\\
      \zeta \hat{G}^{T}&0&0\\
      0&0&-\hat{F}
    \end{array}\right)\, .
  \label{eq:Gmatrixcompact}
\end{equation}
For the last equality, it has been assumed that no symmetry breaking
occurs.  Thus, $\mathbf{G}$ has the same block structure as the
noninteracting $\mathbf{G}_0$ in Eq.~(\ref{eq:G0matrix}) so that,
similarly to Eq.~(\ref{eq:G0transpose}),
\begin{equation}
  \mathbf{G}^T = \mathbf{Z} \mathbf{G}  = \mathbf{G} \mathbf{Z}
  \; .
  \label{eq:Gtranspose}
\end{equation}
In the noninteracting limit ($S_{1} \rightarrow 0$) one easily
verifies by elementary Gaussian integration that the matrix
$\mathbf{G}$ given in Eq.~(\ref{eq:Gmatrix}) reduces to $\mathbf{G}_0$,
as defined in Eq.~(\ref{eq:G0matrix}).  The self-energy matrix also
has the same block structure as the inverse free propagator.  Dyson's
equation then reads as
\begin{equation}
  \mathbf{G}^{-1}=\mathbf{G}_0^{-1}-\mathbf{\Sigma}\,,
  \label{eq:Dyson}
\end{equation}  
where the matrix ${\bf \Sigma}$ contains the one-fermion-line
irreducible self-energy $\Sigma_{\sigma} ( K ) $ and the
one-interaction-line irreducible polarization $\Pi_{\sigma } ( \bar{K}
)$ in the following blocks:
\begin{equation}
    \mathbf{\Sigma} = \left(\begin{array}{ccc}
    0&\zeta [\hat{\Sigma}]^T&0\\
    \hat{\Sigma}&0&0\\
    0&0&\hat{\Pi}
  \end{array}\right)\,,
\end{equation}
where
\begin{eqnarray}
  [\hat{\Sigma} ]_{K\sigma,K'\sigma'}&=& \delta_{K,K'}\delta_{\sigma \sigma'}
  \Sigma_{\sigma} ( K'  ) \,,
  \label{eq:Sigmairdef} \\{}
  [\hat{\Pi}]_{\bar{K}\sigma,\bar{K}'\sigma'}&=& 
  \delta_{\bar{K} + \bar{K}', 0 }  \delta_{\sigma \sigma'}  \,\, \Pi_{\sigma}  
  ( \bar{K}' )  \,.
  \label{eq:Piirdef}
\end{eqnarray}
These matrices are spin-diagonal because the bare coupling $S_1
[\bar{\psi}, \psi , \varphi ]$ between Fermi and Bose fields in
Eq.~(\ref{eq:Spsiphi}) is diagonal in the spin index.  The blocks of
the full Green's function matrix $\mathbf{G}$ in
Eq.~(\ref{eq:Gmatrixcompact}) contain the exact single-particle Green's
function and the effective (screened) interaction,
\begin{eqnarray}
  [\hat{G} ]_{K\sigma,K'\sigma'}&=& \delta_{K,K'}\delta_{\sigma \sigma'}
  G_{\sigma} ( K  ) \,,
  \label{eq:Gfulldef} \\{}
  [\hat{F}]_{\bar{K}\sigma,\bar{K}'\sigma'}&=& 
  \delta_{\bar{K} + \bar{K}', 0 }    \,\,  F_{\sigma \sigma^{\prime}}  
  ( \bar{K} )  \,,
  \label{eq:Ffulldef}
\end{eqnarray}
with
\begin{eqnarray}
  G_{\sigma} ( K ) & = & [ G_{0, \sigma}^{-1} ( K ) - \Sigma_{\sigma} ( K ) ]^{-1}
  \; ,
  \label{eq:Gdiagdef}
  \\
  F_{\sigma \sigma^{\prime}} ( \bar{K} ) & = & 
  \left[  \hat{F}_{0}^{-1}  + \hat{\Pi} \right]^{-1}_{\bar{K} \sigma , -\bar{K} \sigma^{\prime} }
  \; .
  \label{eq:Fdiagdef}
\end{eqnarray}

\subsubsection{Generating functional of one-line irreducible vertices}

Below, we shall derive exact functional RG equations for the one-line
irreducible vertices of our coupled Fermi-Bose theory.  The
diagrammatic perturbation theory consists of both fermion and boson
lines. We require irreducibility with respect to both types of lines
and call this one-line irreducibility.  One should keep in mind that a
boson line represents the two-body electron-electron interaction which
is screened by zero-sound bubbles for small momentum transfers. This
means that in fermionic language our vertices are not only
one-particle irreducible but are also approximately two-particle
irreducible in the zero-sound channel in the sense that particle-hole
bubbles are eliminated in favor of the effective bosonic propagator.
In order to obtain the generating functional of the corresponding
irreducible vertices, we perform a Legendre transformation with
respect to all field components, introducing the classical field
\cite{footnotefield}
\begin{equation}
  \Phi_{\alpha}=\pp{\G_c}{J_{\alpha}}
  \label{eq:def_phi}\,.
\end{equation}
After inverting this relation for $J = J [ \Phi ]$ we may calculate
the Legendre effective action
\begin{equation}
  \LL[\Phi] = (J[\Phi],\Phi)-\G_c[J[\Phi]]\,.
  \label{eq:Ldef}
\end{equation}
From this we obtain
\begin{equation}
  J_{\alpha} = \zeta_{\alpha}\pp{\LL}{\Phi_{\alpha}}\,,
  \label{eq:def_J}
\end{equation}
which we may write in compact matrix notation as
\begin{equation}
  J = \mathbf{Z} \pp{\LL}{\Phi}
  \; .
  \label{eq:def_Jcompact}
\end{equation}
In this notation the chain rule simply reads as
\begin{equation}
  \frac{\delta}{\delta \Phi} =  
  \ppt{\LL}{\Phi}{\Phi}\mathbf{Z}  \frac{\delta}{\delta J }
  \; .
  \label{eq:chain}
\end{equation}
Applying this to both sides of Eq.~(\ref{eq:def_phi}) we obtain
\begin{equation}
  \mathbf{1}=\pp{\Phi}{\Phi}=\ppt{\LL}{\Phi}{\Phi}\mathbf{Z}\ppt{\G_c}{J}{J}\,.
  \label{eq:quad_rel}
\end{equation}
For vanishing fields $\Phi$ and $J$ this yields
\begin{equation}
  \left.\ppt{\LL}{\Phi}{\Phi}\right|_{\Phi=0}=-\mathbf{Z}\mathbf{G}^{-1}
  =- [\mathbf{G}^{-1}]^T
  \; .
  \label{eq:LGcrelation}
\end{equation}
The advantage of our compact notation is now obvious: the minus signs
associated with the Grassmann fields can be neatly collected in the
``statistics matrix'' $\mathbf{Z}$.  If the Grassmann sources are
introduced in the conventional way,\cite{Negele88} the minus signs
generated by commuting two Grassmann fields are distributed in a more
complicated manner in the matrices of
second derivatives.\cite{Kopietz01,Correia01}

From Eq.~(\ref{eq:LGcrelation}) it is evident that we need to subtract
the free action from $\LL [\Phi ]$ to obtain the generating functional
for the irreducible vertex functions,
\begin{equation}
  \Gamma[\Phi]= \LL[\Phi]-S_0[\Phi]=\LL[\Phi]+\frac12(\Phi,[\Go^{-1}]\Phi)\,.
\label{eq:Gammadef}
\end{equation}
Then we have, using the Dyson equation (\ref{eq:Dyson}),
\begin{eqnarray}
  \left.\ppt{\Gamma}{\Phi}{\Phi}\right|_{\Phi=0} &  = &
  \left.\ppt{\LL}{\Phi}{\Phi}\right|_{\Phi=0}
  + [ \mathbf{G}_0^{-1} ]^T
  \nonumber
  \\
  & = & 
  - [ \mathbf{G}^{-1} ]^T + [ \mathbf{G}_0^{-1} ]^T
  =  \mathbf{\Sigma}^T
  \; .
  \label{eq:Sigmarelation}
\end{eqnarray}
In general, the one-line irreducible vertices are defined as
coefficients in an expansion of $\Gamma [ \Phi ]$ with respect to the
fields,
\begin{equation}
  \Gamma[\Phi]=\sum_{n=0}^{\infty}\frac1{n!}\int_{\alpha_1}\dots
  \int_{\alpha_n}\Gamma^{(n)}_{\alpha_1,\dots,\alpha_n}\Phi_{\alpha_1}
  \cdot\ldots\cdot\Phi_{\alpha_n}\,.
  \label{eq:Gammaexpansion}
\end{equation}
The vertices $\Gamma^{(n)}$ have the same symmetry with respect to
interchange of the indices as the monomial in the fields, i.e., the
interchange of two neighboring Fermi fields yields a minus sign.
Graphically, we represent the vertices $\Gamma^{(n)}$
\begin{figure}
  \begin{center}
    \epsfig{file=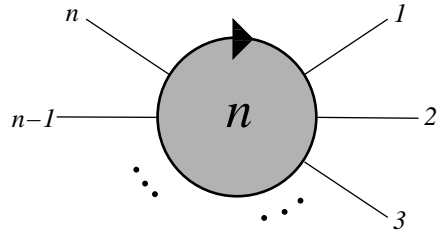,width=0.7\hsize}
  \end{center}
  \caption{Graphical representation of the symmetrized one-line irreducible
    $n$-point vertex defined via Eq.~(\ref{eq:Gammaexpansion}).
    Because for fermions the order of the indices is important
    (exchanging two neighboring fermion legs will generate a minus
    sign) the circles representing the irreducible vertices have an
    arrow that points to the leg corresponding to the first index;
    subsequent indices are arranged in the order indicated by the
    arrow.  External legs denote either outgoing fermions
    ($\bar{\psi}$), incoming fermions ($\psi$), or bosons ($\varphi$).
  }
  \label{fig:generalvertex}
\end{figure}
by an oriented circle with $n$ external legs, as shown in
Fig.~\ref{fig:generalvertex}.  With the
definition~(\ref{eq:Gammaexpansion}) and Eq.~(\ref{eq:Sigmarelation})
we have $\Gamma^{(2)}_{\alpha\alpha'}= [ \mathbf{\Sigma}]_{\alpha'
  \alpha}$.  The fact that also the higher-order vertices
$\Gamma^{(n)}$ defined in Eq.~(\ref{eq:Gammaexpansion}) are indeed
one-line irreducible (i.e., cannot be separated into two parts by
cutting a single fermion line or a single interaction line) can be
shown iteratively by generating a tree expansion from higher-order
derivatives of Eq.~(\ref{eq:quad_rel}).  We show this explicitly in
Appendix~A.

\section{Functional RG flow equations for one-line irreducible vertices}
\label{sec:flow_eq}

In this section we derive exact RG flow equations for the generating
functional of the one-line irreducible vertices of our coupled
Fermi-Bose theory. We also classify the various vertices of the theory
based on their scaling dimensions and propose a new truncation scheme
involving the building blocks of the skeleton diagrams for fermionic
and bosonic two-point functions.

\subsection{Cutoff schemes}

Since the interaction now appears as a propagator of the field
$\varphi$, it is possible to introduce a momentum-transfer cutoff in
the interaction on the same footing as a bandwidth cutoff.  A bandwidth
cutoff restricts the relevant fermionic degrees of freedom to the
vicinity of the Fermi surface, and appears to be most natural in the
RG approach to fermions in one spatial dimension.\cite{Solyom79} In
higher dimensions, the Wilsonian idea of eliminating the degrees of
freedom in the vicinity of the Fermi surface is implemented by
defining for each momentum ${\bf{k}}$ an associated ${\bf{k}}_F$ by
means of a suitable projection onto the Fermi surface \cite{Kopietz01}
and then integrating over fields with momenta in the energy shell $
v_0 \Lambda < | \epsilon_{ {\bf{k}} } - \epsilon_{ {\bf{k}}_F } | <
v_0 \Lambda_0$, where $\epsilon_{\bf{k}} $ is the energy dispersion in
the absence of interactions.  Here $v_0$ is some suitably defined
velocity (for example, some average Fermi velocity), which we introduce
to give $\Lambda$ units of momentum.  We shall refer to $v_0 \Lambda$
as bandwidth cutoff. Formally, we introduce such a cutoff into our
theory by substituting for the free fermionic Green's function in
Eq.~(\ref{eq:G0def})
\begin{eqnarray}
  G_{0, \sigma} (K  ) &\longrightarrow& \Theta( \Lambda< D_K <   \Lambda_0)\,
  G_{0, \sigma} (K )\nonumber\\
  &&= \frac{\Theta(\Lambda< D_K <   \Lambda_0)}{i\omega - \xi_{{\bf k}\sigma}}
  \label{eq:bandcutoff}
  \; ,
\end{eqnarray}
with
\begin{equation}
  D_K  = |\epsilon_{\bf{k}} - \epsilon_{{\bf{k}}_F}| /v_0
  \; .
  \label{eq:OmegaKdef}
\end{equation}
Here $ \Theta ( \Lambda < x < \Lambda_0 ) = 1$ if the logical
expression in the brackets is true, and $ \Theta ( \Lambda < x <
\Lambda_0 ) = 0$ if the logical expression is false.  Ambiguities
associated with the sharp $\Theta$-function cutoff can be avoided by
smoothing out the $\Theta$ functions and taking the sharp cutoff limit
at the end of the calculation.\cite{Morris94}  In order to construct
a consistent scaling theory, the ${\bf{k} }_F$ in
Eq.~(\ref{eq:OmegaKdef}) should refer to the true Fermi surface of the
interacting system, which can in principle be obtained
self-consistently from the condition that the RG flows into a fixed
point.\cite{Kopietz01,Ledowski03}

The above bandwidth-cutoff procedure has several disadvantages.  On
the one hand, for any finite value of the cutoff parameter
$v_0\Lambda$ the Ward identities are violated.\cite{Katanin04}
Moreover, the RG flow of two-particle response functions probing the
response at small momentum transfers (such as the compressibility or
the uniform magnetic susceptibility) is artificially suppressed by the
bandwidth cutoff.  To cure the latter problem, various other
parameters have been proposed to serve as a cutoff for the RG, such as
the temperature \cite{Honerkamp01b} or even the strength of the
interaction.\cite{Honerkamp04} While for practical calculations these
new cutoff schemes may have their advantages, the intuitively
appealing RG picture that the coarse grained parameters of the
renormalized theory contain the effect of the degrees of freedom at
shorter length scales and higher energies gets somewhat blurred (if
not completely lost) by these new schemes.

The above mentioned problems can be elegantly avoided in our mixed
Fermi-Bose theory if we work with a momentum cutoff in the bosonic
sector of our theory, which amounts to replacing in
Eq.~(\ref{eq:F0def}),
\begin{eqnarray}
  F_{0, \sigma \sigma^{\prime} } (\bar{K}) 
  &\longrightarrow& \Theta(\Lambda< \bar{D}_{\bar{K}}
  <\Lambda_0)\,F_{0 , \sigma \sigma^{\prime}} (\bar{K})
  \nonumber\\
  &&=\Theta(\Lambda< \bar{D}_{\bar{K}}
  <\Lambda_0)\,f^{\sigma\sigma^{\prime}}_{\bar{\bf k}}
  \;  ,
  \label{eq:momentumtransfercutoff}
\end{eqnarray}
where
\begin{equation}
  \bar{D}_{\bar{K}} = |\bar{\bf{k}}|
  \; .
  \label{eq:DKbardef}
\end{equation}
Keeping in mind that the bosonic field mediates the effective
interaction, it is clear that $\Lambda$ is a cutoff for the momentum
transfer of the interaction.  This is precisely the same cutoff scheme
employed in the seminal work by Hertz,\cite{Hertz76} who discussed
also more general frequency-dependent cutoffs for the labels of the
bosonic Hubbard-Stratonovich fields, corresponding to more complicated
functions $\bar{D}_{\bar{K}}$ than the one given in
Eq.~(\ref{eq:DKbardef}).  Moreover, in the exact solution of the
one-dimensional Tomonaga-Luttinger model (abbreviated here as TLM, as
already defined above) by means of a careful application of the
bosonization method \cite{Schoenhammer03} the maximal momentum
transfered by the interaction appears as the natural cutoff scale.

In our RG approach we have the freedom of choosing both the bandwidth
cutoff $v_0 \Lambda $ and the momentum-transfer cutoff $\Lambda$
independently.  In particular, we may even choose to get rid of the
bandwidth cutoff completely and work with a momentum-transfer cutoff
only.  In this work, we shall show that if the interaction involves
only small momentum transfers, then the pure momentum-transfer cutoff
scheme indeed regularizes all infrared singularities in one dimension.
Moreover and most importantly, introducing a cutoff only in the
momentum transfer leads to exact RG flow equations that do not
violate the Ward identities responsible for the exact solubility of
the TLM.  Given this fact, it is not surprising that we can solve the
infinite hierarchy of RG flow equations exactly and obtain the exact
single-particle Green's function of the TLM within the framework of the
functional RG.

\subsection{Flow equations for  completely symmetrized vertices}

With the substitutions (\ref{eq:bandcutoff}) and
(\ref{eq:momentumtransfercutoff}) the noninteracting Green's function
$\mathbf{G}_0$, and hence all generating functionals, depend on the
cutoff parameter $\Lambda$.  We can now follow the evolution of the
generating functionals as we change the cutoff.  The differentiation of
Eq.~(\ref{eq:Ggen}) with respect to $\Lambda$ yields for the
generating functional of the Green's functions,
\begin{equation}
  \partial_{\Lambda}\G = \left\{
    \frac12\left(
      \p{J},\partial_{\Lambda}[\Go^{-1}]\p{J}\right)
    -\partial_{\Lambda}\ln\Z_0
  \right\}\G\,.
\end{equation}
For the connected version $\G_c [ J ] = \ln \G [ J ]$, we obtain
\begin{eqnarray}
  \partial_{\Lambda}\G_c = 
  \frac12\left(
    \pp{\G_c}{J},\partial_{\Lambda}[\Go^{-1}]\pp{\G_c}{J}\right)
  ~~~~~~~~~~~~~~~~~~~~~~~&&\nonumber\\
  +\frac12\Tr\left(\partial_{\Lambda}[\Go^{-1}]\left[\ppt{\G_c}{J}{J}\right]^T\right)
  -\partial_{\Lambda}\ln\Z_0\,.&&
  \label{eq:Gcflow}
\end{eqnarray}
In the derivation of flow equations for $\LL$ or $\Gamma$ [see
Eqs.~(\ref{eq:Ldef}) and (\ref{eq:Gammadef})], we should keep in mind
that in these functionals the fields $\Phi$ are held constant rather
than the sources $J$. Hence, Eq.~(\ref{eq:Ldef}) implies
\begin{equation}
  \partial_{\Lambda}\LL[\Phi]=-\left.\partial_{\Lambda}\G_c[J]\right|_{J=J_{\Lambda}[\Phi]}
  \; .
\end{equation}
Using this and Eq.~(\ref{eq:Gcflow}) we obtain for the functional
$\Gamma [\Phi ] = \LL [\Phi] - S_0 [\Phi]$,
\begin{eqnarray}
  \partial_{\Lambda} \Gamma = 
  & - & \frac12\Tr\left(\partial_{\Lambda}[\Go^{-1}]\left[\ppt{\G_c}{J}{J}\right]^T\right)
  + \partial_{\Lambda}\ln\Z_0 \; .
  \nonumber
  \\
  & &
  \label{eq:Gammaflowprelim}
\end{eqnarray}
To derive a closed equation for $\Gamma$, we express the matrix
$\ppt{\G_c}{J}{J}$ in terms of derivatives of $\Gamma$ using
Eq.~(\ref{eq:GcJJexpansion}).  After some rearrangements we obtain the
exact flow equation for the generating functional $\Gamma [\Phi ] $ of
the one-line irreducible vertices,
\begin{eqnarray}
  \partial_{\Lambda}\Gamma&=&
  -\frac12\Tr\left[\mathbf{Z}\dot{\mathbf{G}}^T\mathbf{U}^T
    \left\{
      \mathbf{1}-\mathbf{G}^T\mathbf{U}^T
    \right\}^{-1}\right]\nonumber\\[0.2cm]
  &&-\frac12\Tr\left[\mathbf{Z}\dot{\mathbf{G}}_0^T\mathbf{\Sigma}^T
    \left\{
      \mathbf{1}-\mathbf{G}_0^T\mathbf{\Sigma}^T
    \right\}^{-1}\right]\,,
  \label{eq:Gammaflow}
\end{eqnarray}
\begin{figure}[t]
  \begin{center}
    \epsfig{file=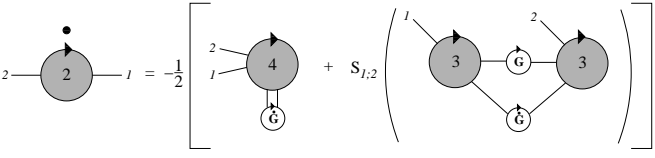,width=1.0\hsize}
  \end{center}
  \caption{Graphical representation of Eq.~(\ref{eq:flow_vert}) for $n=2$, describing
    the flow of the totally symmetric two-point vertex.  Empty circles
    with ${\bf G}$ and $\dot{\bf G}$ denote the exact matrix propagator
    ${\bf G}$ and the single-scale propagator $\dot{\bf G}$ defined in
    Eqs.~(\ref{eq:Gmatrix}) and (\ref{eq:Gdotmatrix}), respectively.}
  \label{fig:flowGamma2}
\end{figure}%
where the matrix $\mathbf{U} [ \Phi ] $ is the field-dependent part of
the second functional derivative of $\Gamma [\Phi]$, as defined in
Eq.~(\ref{eq:Udef}).  For convenience we have introduced the 
single-scale propagator $\dot{\mathbf{G}}$ as
\begin{equation}
  \dot{\mathbf{G}} = - \mathbf{G}\partial_{\Lambda}[\mathbf{G}_0^{-1}]\mathbf{G}
  =[\mathbf{1}-\mathbf{G}_0\mathbf{\Sigma}]^{-1}
  \left(\partial_{\Lambda}\mathbf{G}_0
  \right)
  [\mathbf{1}-\mathbf{\Sigma}\mathbf{G}_0]^{-1}
  \; ,
  \label{eq:Gdotmatrix}
\end{equation}%
which reduces to $\dot{\mathbf{G}}_0 = \partial_{\Lambda}
\mathbf{G}_0$ in the absence of interactions.  The matrix
$\dot{\mathbf{G}}$ has the same block structure as the matrix
$\mathbf{G}$ in Eq.~(\ref{eq:Gmatrixcompact}).  We denote the
corresponding blocks by $\dot{\hat{G}}$ and $\dot{\hat{F}}$.  In the
limit of a sharp $\Theta$-function cutoff~\cite{Morris94} the blocks
of the single scale propagator are explicitly given by
\begin{eqnarray}
  [\dot{\hat{G} }]_{K\sigma,K'\sigma'}&=& \delta_{K,K'}\delta_{\sigma \sigma'}
  \dot{G}_{\sigma} ( K  ) \,,
  \label{eq:Gsinglescale} \\{}
  [\dot{\hat{F}}]_{\bar{K}\sigma,\bar{K}'\sigma'}&=& 
  \delta_{\bar{K} + \bar{K}', 0 }    \,\,  \dot{F}_{\sigma \sigma^{\prime}}  
  ( \bar{K} )  \,,
  \label{eq:Fsinglescale}
\end{eqnarray}
with
\begin{eqnarray}
  \hspace{-4mm} \dot{G}_{\sigma} ( K ) & = & - \frac{ \delta ( \Lambda - D_K ) }{i \omega
    - \xi_{ {\bf{k}} \sigma} - \Sigma_{\sigma} ( K ) }
  \; ,
  \label{eq:dotGdiagdef}
  \\
  \hspace{-4mm}  \dot{F}_{\sigma \sigma^{\prime}} ( \bar{K} ) & = &  -
  \delta ( \Lambda - \bar{D}_{\bar{K}} )
  \left[  \hat{F}_{0}^{-1}  + \hat{\Pi} \right]^{-1}_{ \bar{K} \sigma , -\bar{K} \sigma^{\prime} }
  \; ,
  \label{eq:dotFdiagdef}
\end{eqnarray}
where on the right-hand side of Eq.~(\ref{eq:dotFdiagdef}) it is
understood that the $\Theta$-function cutoff should be omitted from
the matrix elements of $\hat{F}_0$.

The second line in Eq.~(\ref{eq:Gammaflow}) does not depend on the
fields any longer and therefore represents the flow of the interaction
correction  $\Gamma^{(0)}$ to the grand-canonical potential,
\begin{equation}
  \partial_{\Lambda} \Gamma^{(0)} = 
  -\frac12\Tr\left[\mathbf{Z}\dot{\mathbf{G}}_0^T\mathbf{\Sigma}^T
    \left\{
      \mathbf{1}-\mathbf{G}_0^T\mathbf{\Sigma}^T
    \right\}^{-1}\right]\,.
  \label{eq:Gamma0flow}
\end{equation}
Since we have already dropped constant parts of the action in the
Hubbard-Stratonovich transformation, we will not keep track of
$\Gamma^{(0)}$ in the following.  

The first line on the right-hand side of Eq.~(\ref{eq:Gammaflow})
gives the flow of one-line irreducible vertices.  We can generate a
hierarchy of flow equations for the vertices by expanding both sides
in powers of the fields. On the left-hand side, we simply insert the
functional Taylor expansion (\ref{eq:Gammaexpansion}) of $\Gamma [
\Phi ]$, while on the right-hand side we substitute the expansion of
$\mathbf{U} [ \Phi ]$ given in Eq.~(\ref{eq:Uexpansion}).  For a
comparison of the coefficients on both sides, the right-hand side has
to be symmetrized with respect to external lines on different
vertices.  We can write down the resulting infinite system of flow
equations for the one-line irreducible vertices $\Gamma^{(n)}$ with $n
\geq 1$ in the following closed form:
\begin{widetext}
  \begin{eqnarray} \partial_{\Lambda} {\Gamma}^{(n)}_{\alpha_1,\dots,\alpha_n} =
    -\frac12\sum\limits_{l=1}^{\infty}
    \sum\limits_{m_1,\dots,m_l=1}^{\infty}\delta_{n,m_1+\ldots+m_l}\,
    {\cal{S}}_{\alpha_1,\dots,\alpha_{m_1};\alpha_{m_1+1},\dots,\alpha_{m_1+m_2};\dots;\alpha_{m_1+\ldots+m_{l-1}+1},\dots,\alpha_{n}}
    \Big\{~~~~~~~~&&\nonumber\\   \times \Tr\left[
      \mathbf{Z}\dot{\mathbf{G}}^T \mathbf{\Gamma}^{(m_1+2)\,T}_{\alpha_1,\dots,\alpha_{m_1}}\mathbf{G}^T
      \mathbf{\Gamma}^{(m_2+2)\,T}_{\alpha_{m_1+1},\dots,\alpha_{m_1+m_2}}\mathbf{G}^T
      \dots
      \mathbf{\Gamma}^{(m_l+2)\,T}_{\alpha_{m_1+\ldots+m_{l-1}+1},\dots,\alpha_{n}}
    \right] \Big\}\,.
    \label{eq:flow_vert}  
  \end{eqnarray}
\end{widetext}
Here the matrices $\mathbf{\Gamma}^{(m+2)}_{\alpha_1 \ldots \alpha_m}$
are given in Eq.~(\ref{eq:Gammamatrix}) and the symmetrization
operator ${\cal{S}}$ is defined in Eq.~(\ref{eq:symmopdef}).  The
effect of ${\cal{S}}$ is rather simple: it acts on an expression
already symmetric in the index groups separated by semicolons to
generate an expression symmetric also with respect to the exchange of
indices between different groups.  From one summand in
Eq.~(\ref{eq:flow_vert}) the symmetrization operator ${\cal{S}}$ thus
creates $n!/(m_1!\cdot\ldots\cdot m_l!)$ terms.
\begin{figure}[t]
  \begin{center}
    \epsfig{file=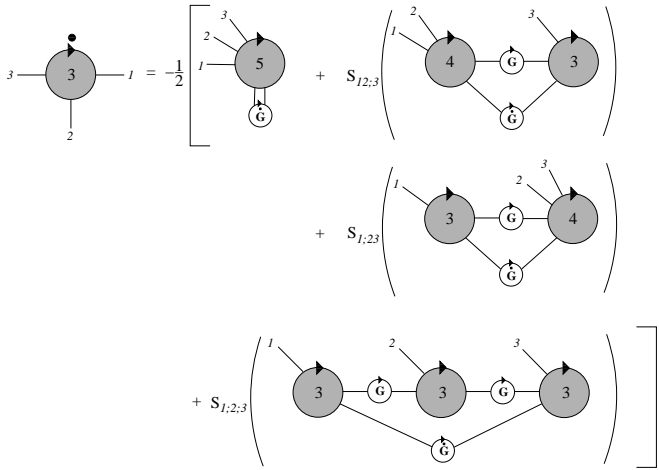,width=1.0\hsize}
  \end{center}
  \caption{Graphical representation of Eq.~(\ref{eq:flow_vert}) for $n=3$, describing
    the flow of the totally symmetric three-point vertex.}
\label{fig:flowGamma3}
\end{figure}

Figures.~\ref{fig:flowGamma2} and \ref{fig:flowGamma3} show a graphical
representation of the flow of the vertices $\Gamma^{(2)}$ and
$\Gamma^{(3)}$.  With the graphical notation for the totally symmetric
vertices introduced in Fig.~\ref{fig:generalvertex} all the signs and
combinatorics have a graphical representation.  In the next section
we will leave the shorthand notation and go back to more physical
vertices, explicitly exhibiting the different types of fields.  All
this can be done on a graphical level and involves only
straightforward combinatorics. In this sense the derivation of higher
flow equations is at the same level of complexity as ordinary Feynman
graph expansions.

\subsection{Flow equations for physical vertices}
\label{subsec:flowphysical}

\begin{figure}[b]
  \begin{center} 
    \epsfig{file=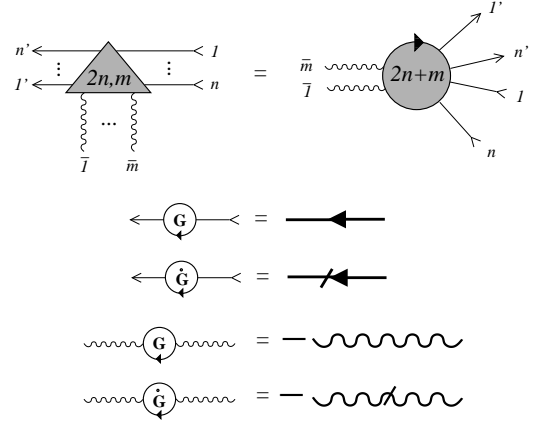,width=0.8\hsize}
    \vspace{0.2cm}
    \caption{Pictorial
      dictionary to translate graphs involving totally symmetrized
      vertices to ones involving physical vertices, which are only
      symmetrized within fields of the same type.  The diagrams on the
      right-hand sides of the last four lines represent $G$,
      $\dot{G}$, $F$ and $\dot{F}$ respectively. }
    \label{fig:picdic}
    \end{center}
\end{figure}
\begin{figure}
  \begin{center}
    \epsfig{file=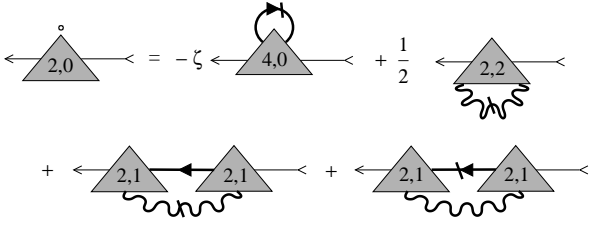,width=0.9\hsize}
  \end{center}
  \caption{Flow of the irreducible fermionic self-energy.
    The diagrams are obtained from the diagrams shown in
    Fig.~\ref{fig:flowGamma2} by specifying the external legs to be
    one outgoing and one incoming fermion leg.}
  \label{fig:flowSigma}
\end{figure}
\begin{figure}
  \begin{center}
    \epsfig{file=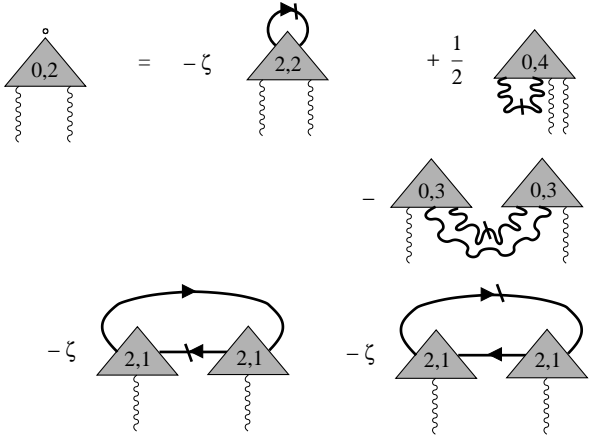,width=0.9\hsize}
  \end{center}
  \caption{Flow of the irreducible polarization, obtained from the
    totally symmetric diagram in Fig.~\ref{fig:flowGamma2} by setting
    both external legs equal to boson legs. Note that each closed
    fermion loop gives rise to an additional factor of $\zeta=-1$.}
  \label{fig:flowPi}
\end{figure}

Usually, the generating functional $\Gamma [ \bar{\psi}, \psi ,
\varphi ]$ is expanded in terms of correlation functions that are not
symmetrized with respect to the exchange of legs involving different
types of fields. If we explicitly display momentum and frequency
conservation, such an expansion reads as
\begin{widetext}
  \begin{eqnarray}
    \Gamma[\bar{\psi}, \psi,\varphi]  & =& \sum_{n=0}^{\infty}\sum_{m=0}^{\infty}
    \frac{1}{(n!)^2m!}
    \int_{K_1'\sigma_1'}\dots\int_{K_n'\sigma_n'}
    \int_{K_1\sigma_1}\dots\int_{K_n\sigma_n}
    \int_{\bar{K}_1\bar{\sigma}_1}\dots\int_{\bar{K}_m\bar{\sigma}_m}
    \delta_{ K'_1 + \ldots + K'_n ,  K_1 + \ldots + K_n  + \bar{K}_1 + \ldots + \bar{K}_m }
    \nonumber \\
    & & \hspace{20mm}  \times
    \Gamma^{(2n,m)}(K'_1\sigma_1',\dots,K'_n\sigma_n';
    K_1\sigma_1,\dots,
    K_n\sigma_n;\bar{K}_1\bar{\sigma}_1,\dots,\bar{K}_m\bar{\sigma}_m) \nonumber\\
    & & \hspace{20mm} \times
    \bar{\psi}_{K'_1\sigma_1'}\cdot\ldots\cdot\bar{\psi}_{K'_n\sigma_n'}
    \psi_{K_1\sigma_1}\cdot\ldots\cdot\psi_{K_n\sigma_n}
    \varphi_{\bar{K}_1\bar{\sigma}_1}\cdot\ldots\cdot\varphi_{\bar{K}_m\bar{\sigma}_m}\,.
    \label{eq:expansion2}
  \end{eqnarray}
\end{widetext}
Diagrammatically, we represent a physical vertex $\Gamma^{(2n,m)}$
involving $2n$ external fermion legs and $m$ external boson legs by a
triangle to emphasize that our theory contains three types of fields,
see Fig.~\ref{fig:picdic}.
We represent a leg associated with a $\bar{\psi}$ field by an arrow
pointing outward, a leg for $\psi$ by an arrow pointing inward, and
a leg for $\varphi$ with a wiggly line without an arrow.  Recall that
our Bose field is real because it couples to the density, so that it
should be represented graphically by an undirected line.  On the
contrary, for a propagator $\mathbf{G}$ or $\dot{\mathbf{G}}$, the
field $\bar{\psi}$ is represented by an arrow pointing inward and
$\psi$ by an arrow pointing outward.  Apart from the energy- and
momentum-conserving delta function, the totally symmetric vertices
defined by the expansion (\ref{eq:Gammaexpansion}) coincide with the
nonsymmetric ones in Eq.~(\ref{eq:expansion2}) for the same order of
the indices.  We can therefore obtain the flow equations for the
nonsymmetric vertices by choosing a definite realization of the
external legs and by carrying out the intermediate sums over the
different field species, i.e., by drawing all possible lines in the
intermediate loop (two possible orientations of solid lines or one
wiggly line). On the right-hand side one then has to appropriately
order all the legs on the vertices, keeping track of signs for the
interchange of two neighboring fermion legs. Having done so, we can
use the pictorial dictionary in Fig.~\ref{fig:picdic} to obtain
diagrams involving the physical correlation functions.  In this way,
we obtain from the diagram for the completely symmetric two-point
vertex shown in Fig.~\ref{fig:flowGamma2} the diagram for the
fermionic self-energy in Fig.~\ref{fig:flowSigma} as well as the
diagram for the irreducible polarization shown in
Fig.~\ref{fig:flowPi}.
\begin{figure}
  \begin{center}
    \epsfig{file=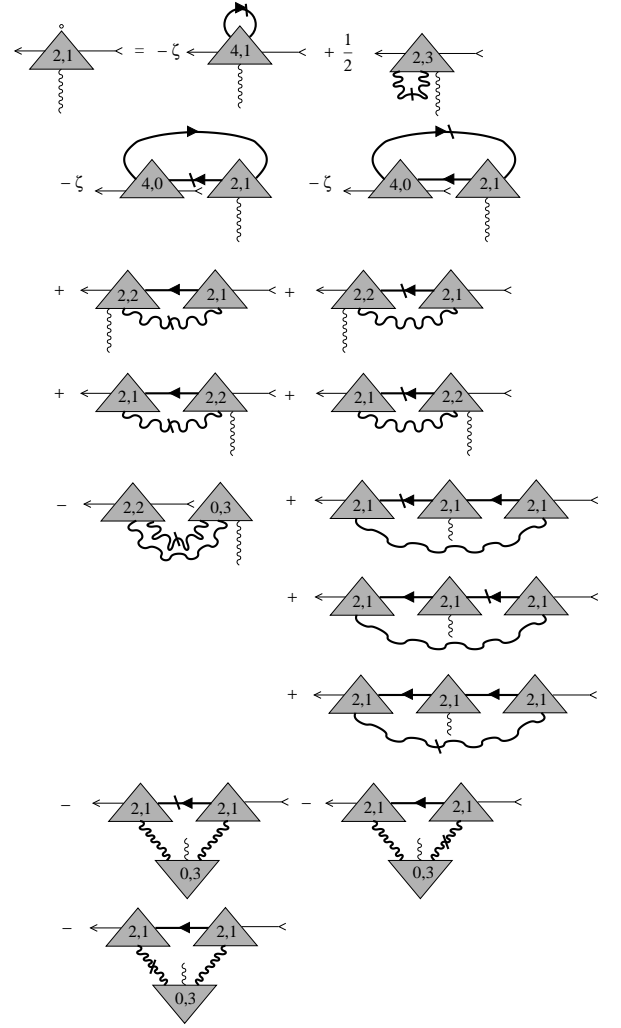,width=0.9\hsize}
  \end{center}
  \caption{Flow of three-legged vertex with two fermion legs and one boson leg, obtained as a special case
    of the diagram in Fig.~\ref{fig:flowGamma3}.  }
  \label{fig:flowVCorr}
\end{figure} 
Moreover, if we specify the external legs in the diagram for the
completely symmetric three-legged vertex shown in
Fig.~\ref{fig:flowGamma3} to be two fermion legs and one boson leg, we
obtain the flow equation for the three-legged vertex shown in
Fig.~\ref{fig:flowVCorr}.

The flow equation for the vertex correction in
Fig.~\ref{fig:flowVCorr} looks very complicated, so that at this point
the reader might wonder how in one dimension we will be able to obtain
the exact solution of the TLM using our approach.  We shall explain
this in detail in Sec.~\ref{sec:1dflow}, but let us anticipate here
the crucial step: obviously all diagrams shown in
Figs.~\ref{fig:flowSigma}, \ref{fig:flowPi}, and \ref{fig:flowVCorr}
can be subdivided into two classes: those involving a fermionic
single-scale propagator (the slash appears on an internal fermion
line), and those with a bosonic single-scale propagator (with a slash
on an internal boson line).  At this point we have not specified the
cutoff procedure, but as mentioned in the Introduction, in
Sec.~\ref{sec:1dflow} we shall work with a bosonic cutoff only. In
this case all diagrams with a slash attached to a fermion line should
simply be omitted. This is the crucial simplification which will allow
us to solve the hierarchy of flow equations exactly.  Let us proceed
in this section without specifying a particular cutoff procedure.

\subsection{Rescaling and classification of vertices}
\label{subsec:rescale}

In order to assign scaling dimensions to the vertices, we have to
define how we rescale momenta, frequencies, and fields under the RG
transformation.  The rescaling is not unique but depends on the nature
of the fixed point we are looking for. Let us be general here and
assume that in the bosonic sector the relation between momentum and
frequency is characterized by a bosonic dynamic exponent $z_\varphi$
(this is the exponent $z$ introduced by Hertz\cite{Hertz76}), while
in the fermionic sector the corresponding dynamic exponent is
$z_{\psi}$.  Rescaled dimensionless bosonic momenta $\bar{\bf{q}}$ and
frequencies $\bar{\epsilon}$ are then introduced as
usual\cite{Hertz76}
\begin{equation}
  \bar{\bf{q}} = \bar{\bf{k}} / \Lambda \; \; , \; \; \bar{\epsilon} = \bar{\omega} / 
  \bar{\Omega}_{\Lambda} \; \; ,
  \; \; \bar{\Omega}_{\Lambda}  \propto \Lambda^{z_{\varphi}}
  \; .
  \label{eq:bosrescale}
\end{equation}
For convenience, we choose the factor $\bar{\Omega}_{\Lambda}$ such
that it has units of energy; $\bar{\epsilon}$ is then dimensionless.

The proper rescaling of the fermionic momenta is not so obvious.
Certainly, all momenta should be measured with respect to suitable
points ${\bf{k}}_F$ on the Fermi surface. One possibility is to
rescale only the component $k_{\parallel} = ( {\bf{k}} - {\bf{k}}_F )
\cdot \hat{\bf{v}}_F$ of a given momentum that is parallel to the local
Fermi surface velocity ${\bf{v}}_F$ (and hence perpendicular to the Fermi
surface),\cite{Shankar94,Kopietz01}  where $\hat{\bf v}_F$ is a unit vector 
in the direction of ${\bf v}_F$. Unfortunately, in dimensions $D >
1$ this leads to rather complicated geometric constructions, because
in a fixed reference frame the component $k_{\parallel}$ to be
rescaled varies for different points on the Fermi surface.  However,
if the initial momentum-transfer cutoff $\Lambda_0$ in
Eq.~(\ref{eq:momentumtransfercutoff}) is small compared with the
typical radius of the Fermi surface, the initial and final momenta
associated with a scattering process lie both on nearby points on the
Fermi surface.  In this case it seems natural to pick one fixed
reference point ${\bf{k}}_{F, \sigma}$ on the Fermi surface, and then
measure all fermionic momentum labels ${\bf{k}}_i$ and
${\bf{k}}_i^{\prime}$ in $\Gamma^{(2n,m)}(K'_1,\dots,K'_n; K_1,\dots,
K_n;\bar{K}_1,\dots,\bar{K}_m) $ relative to this ${\bf{k}}_{F,
  \sigma}$.  Here, the index $\sigma$ labels the different points on
the Fermi surface, for example, in one dimension $\sigma = \pm 1$, with
$k_{F,\pm 1} = \pm k_F$.  We then define rescaled fermionic momenta
${\bf{q}}$ and frequencies ${\epsilon}$ as follows:
\begin{equation}
  {\bf{q}} = ( {\bf{k}}  - {\bf{k}}_{F, \sigma} )/ \Lambda \; \; , \; \; {\epsilon} = {\omega} / 
  \Omega_{\Lambda} \; \; ,
  \; \; \Omega_{\Lambda}  \propto \Lambda^{z_{\psi}}
  \; .
  \label{eq:fermirescale}
\end{equation}
The factor $\Omega_{\Lambda}$ should again have units of energy such
that $\epsilon$ is dimensionless.  Iterating the usual RG steps
consisting of mode elimination and rescaling, we then coarse grain the
degrees of freedom in a sphere around the chosen point ${\bf{k}}_{F ,
  \sigma}$.  Because by assumption the maximal momentum transfer
mediated by the interaction is small compared with $ | {\bf{k}}_{F,
  \sigma} |$, the fermionic momenta appearing in
$\Gamma^{(2n,m)}(K'_1,\dots,K'_n; K_1,\dots,
K_n;\bar{K}_1,\dots,\bar{K}_m) $ are all in the vicinity of the chosen
${\bf{k}}_{F , \sigma}$.  This property is also responsible for the
approximate validity of the {\it{closed loop theorem}} for interacting
fermions with dominant forward scattering in arbitrary
dimensions.\cite{Kopietz95,Kopietz97,Metzner98}

Apart from the rescaling of momenta and frequencies, we have to
specify the rescaling of the fields. As usual, we require that the
Gaussian part $S_0 [ \Phi ] = S_0 [ \bar\psi , \psi ] + S_0 [\varphi]$
of our effective action is invariant under rescaling.  For the
fermionic part this is achieved by defining renormalized fields
$\tilde{\psi}_{Q\sigma}$ in $D$ dimensions via
\begin{equation}
  \psi_{K \sigma}  =
  \left(  
    \frac{ Z }{ \Lambda^D \Omega_{\Lambda}^2 }  \right)^{1/2} \tilde{\psi}_{ Q \sigma } 
  \; ,
  \label{eq:Fermifieldrescale}
\end{equation}
where $Z$ is the fermionic wave-function renormalization factor and $Q
= ( {\bf{q}} , i \epsilon )$ denotes the rescaled fermionic momenta
and Matsubara frequencies as defined in Eq.~(\ref{eq:fermirescale}).
With this rescaling the wave-function renormalization and the Fermi
velocity have a vanishing scaling dimension (corresponding to marginal
couplings), while the momentum- and frequency-independent part of the
self-energy is relevant with scaling dimension $+1$; see
Ref.~\onlinecite{Kopietz01}.  Analogously, we find that the bosonic Gaussian
part of the action is invariant under rescaling if we express it in
terms of the renormalized bosonic field $\tilde{\varphi}_{\bar{Q}
  \sigma }$ defined by
\begin{equation}
  \varphi_{\bar{K} \sigma}  =
  \left(  
    \frac{ \bar{Z} }{ \Lambda^D \bar{\Omega}_{\Lambda} \nu_0 }  \right)^{1/2} 
  \tilde{\varphi}_{ \bar{Q} \sigma } 
  \; ,
  \label{eq:Bosefieldrescale}
\end{equation}
where $\bar{Z}$ is the bosonic wave-function renormalization factor,
$\bar{Q} = ( \bar{\bf{q}} , i \bar{\epsilon} )$ denotes the rescaled
bosonic momenta and Matsubara frequencies defined in
Eq.~(\ref{eq:bosrescale}), and $\nu_0$ denotes the noninteracting
density of states at the Fermi surface.  We introduce the factor of
$\nu_0$ for convenience to make all rescaled vertices dimensionless.
By construction Eq.~(\ref{eq:Bosefieldrescale}) assigns vanishing
scaling dimensions to the bare interaction parameters $ f_{
  \bar{\bf{k}} }^{\sigma \sigma^{\prime} }$, corresponding to marginal
Landau interaction parameters.

Expressing each term in the expansion of the generating functional
$\Gamma [ \bar{\psi} , \psi , \varphi ]$ given in
Eq.~(\ref{eq:expansion2}) in terms of the rescaled variables defined
above and using the fact that $\Gamma$ is dimensionless, we obtain the
scaling form of the vertices.  Omitting for simplicity the degeneracy
labels $\sigma$, and assuming $z_{\psi}\leq z_{\varphi}$,\cite{zfootnote}
we define the rescaled vertices,
\begin{eqnarray}
  \tilde{\Gamma}_l^{(2n,m)}(Q_1',\dots,Q_n';Q_1,\dots,Q_n;\bar{Q}_1,\dots,\bar{Q}_m)
  = &  &
  \nonumber\\
  & & \hspace{-76mm}  
  \nu_0^{-m/2} \Lambda^{D ( n -1 + m/2) } 
  \Omega_{\Lambda}^{-1} \bar{\Omega}_{\Lambda}^{{m}/{2}}
  Z^n \bar{Z}^{{m}/2}
  \nonumber
  \\
  & & \hspace{-79mm} \times
  \Gamma_{\Lambda}^{(2n,m)}(K_1',\dots,K_n';K_1,\dots,K_n;\bar{K}_1,\dots,\bar{K}_m)\,,
  \label{eq:Gammarescaledef}
\end{eqnarray}
where we have to exclude the cases of purely bosonic vertices ($n=0$)
as well as the fermionic two-point vertex (i.e., the rescaled
irreducible self-energy, corresponding to $n=1$ and $m=0$), which both
need separate definitions. For the purely bosonic vertices ($n=0$) we
set
\begin{eqnarray}
  \tilde{\Gamma}^{(0,m)}_{l} ( \bar{Q}_1,\dots,\bar{Q}_m   ) & = &
  \nu_0^{-m/2} (  \Lambda^{D} \bar{\Omega}_{\Lambda} )^{ -1 + {m}/{2} } 
  \bar{Z}^{{m}/2}
  \nonumber
  \\
  & \times &
  \Gamma_{ \Lambda }^{(0,m)}  (  \bar{K}_1,\dots,\bar{K}_m) 
  \; ,
  \label{eq:Gammarescaledefbos}
\end{eqnarray}
while for the fermionic two-point vertex we should subtract the exact
fixed point self-energy $ \Sigma_{\ast} ( {\bf{k}}_{F ,\sigma} , i0 )$
at the Fermi-surface reference-point ${\bf{k}}_{F, \sigma}$ and for
vanishing frequency as a counterterm,\cite{Kopietz01,Ledowski03}
\begin{equation}
  \tilde{\Gamma}^{(2,0)}_l ( Q ; Q )
  \equiv \tilde{\Sigma}_l (Q ) = \frac{Z}{\Omega_{\Lambda}}
  \left[ \Sigma ( K ) - \Sigma_{\ast} ( {\bf{k}}_{F, \sigma} , i0 ) \right]
  \; .
  \label{eq:Sigmarescaledef}
\end{equation} 
If necessary, the counterterm $ \Sigma_{\ast} ( {\bf{k}}_{F ,\sigma}
, i0 )$ can be reconstructed from the condition that the constant part
$\tilde{r}_l = \tilde{\Sigma}_l ( 0 )$ of the self-energy flows into
an RG fixed point.\cite{Kopietz01,Ledowski03}  We consider the
rescaled vertices to be functions of the logarithmic flow parameter $l
= - \ln ( \Lambda / \Lambda_0 )$.  Introducing the flowing anomalous
dimensions associated with the fermionic and bosonic fields,
\begin{equation}
  \eta_l = -\partial_l \ln Z \; \; \; , \; \; \; 
  \bar{\eta}_l = -\partial_l \ln \bar{Z} \,,
  \label{eq:etadef}
\end{equation}
we can then write down the flow equations for the rescaled vertices.
Omitting the arguments, we obtain for $n \geq 1$ the flow equation\cite{zfootnote}
\begin{widetext}
  \begin{eqnarray}
    \partial_l \tilde{\Gamma}^{(2n,m)}_l  &=&
    \left[   (1-n) D + z_{\text{min}} - \frac{m}{2} ( D + z_\varphi ) -  n \eta_l -\frac{m}2 \bar{\eta}_l 
      - \sum_{i=1}^n (Q_i' \cdot  \frac{\partial}  {\partial Q_i'} + 
      Q_i \cdot \frac{\partial}{\partial Q_i})
      - \sum_{i=1}^m \bar{Q}_i \cdot \frac{\partial}{\partial \bar{Q}_i}
    \right] 
    \tilde{\Gamma}_l^{(2n,m)}
    \nonumber\\
    &&+\;\dot{\tilde{\Gamma}}_l^{(2n,m)}\,,
    \label{eq:flowGammarescale1}
  \end{eqnarray}
  where $z_{\text{min}}=\text{min}\{z_{\varphi},z_{\psi}\}$. For $n=0$
  we obtain from Eq.~(\ref{eq:Gammarescaledefbos}),
  \begin{equation}
    \partial_l \tilde{\Gamma}^{(0,m)}_l =
    \left[   (1 - \frac{m}{2})  ( D + z_\varphi ) -\frac{m}2 \bar{\eta}_l 
      - \sum_{i=1}^m \bar{Q}_i \cdot \frac{\partial}{\partial \bar{Q}_i}
    \right] 
    \tilde{\Gamma}_l^{(0,m)}
    \,+\,\dot{\tilde{\Gamma}}_l^{(0,m)}\,,
    \label{eq:flowGammarescale2}
  \end{equation}
\end{widetext}
where we have introduced the notation
\begin{eqnarray}
  Q \cdot \frac{\partial }{ \partial Q}  & \equiv &  {\bf{q}} \cdot \mathbf{\nabla}_{\bf{q}} + z_{\psi} \; \epsilon
  \frac{\partial}{\partial \epsilon}
  \label{eq:QdQdef}
  \; ,
  \\
  \bar{Q} \cdot \frac{\partial }{ \partial \bar{Q}}  & \equiv &  \bar{\bf{q}} \cdot \mathbf{\nabla}_{\bar{\bf{q}}} 
  + z_{\varphi} \; \bar{\epsilon}
  \frac{\partial}{\partial \bar{\epsilon}}
  \label{eq:barQdQdef}
  \; .
\end{eqnarray}
The inhomogeneities in Eqs.~(\ref{eq:flowGammarescale1}) and
(\ref{eq:flowGammarescale2}) are given by the rescaled version of the
right-hand sides of the flow equations for the unrescaled vertices,
i.e., for $n \geq 1$, and $z_{\psi}\leq z_{\varphi}$,\cite{zfootnote}
\begin{eqnarray}
  \dot{\tilde{\Gamma}}_l^{(2n,m)}(Q_1',\dots,Q_n';Q_1,\dots,Q_n';\bar{Q}_1,\dots,\bar{Q}_m)
  & = &
  \nonumber
  \\
  & & \hspace{-70mm}  
  \nu_0^{-m/2} \Lambda^{D ( n -1 + m/2) } 
  \Omega_{\Lambda}^{-1} \bar{\Omega}_{\Lambda}^{{m}/{2}}
  Z^n \bar{Z}^{{m}/2}
  \nonumber
  \\
  & & \hspace{-70mm} \times
  [ - \Lambda \partial_{\Lambda} \Gamma_{\Lambda}^{(2n,m)}( \left\{ K_i' ;K_i;\bar{K}_i\right\}) ]
  \label{eq:inhrescale1}
  \; ,
\end{eqnarray}
and for $n=0$,
\begin{eqnarray}
  \dot{\tilde{\Gamma}}^{(0,m)}_{l} ( \bar{Q}_1,\dots,\bar{Q}_m   ) & = &
  \nu_0^{-m/2} (  \Lambda^{D} \bar{\Omega}_{\Lambda} )^{ -1 + {m}/{2} } 
  \bar{Z}^{{m}/2}
  \nonumber
  \\
  &  & \hspace{-15mm} \times
  [ - \Lambda \partial_{\Lambda}
  \Gamma_{ \Lambda }^{(0,m)}  (  \bar{K}_1,\dots,\bar{K}_m) ] \, . 
  \label{eq:inhrescale2}
\end{eqnarray}
By properly counting all factors it is then not difficult to see that
the explicit expressions for the inhomogeneities in
Eqs.~(\ref{eq:inhrescale1}) and (\ref{eq:inhrescale2}) can be simply
obtained from their unrescaled counterparts by replacing all vertices
and propagators with their rescaled analogs, where the rescaled
propagators are defined by
\begin{equation}
  {G}(K)=\frac{Z}{\Omega_{\Lambda} } \tilde{G}(Q)\,,
  \; \; \;
  {F}(\bar{K})=\frac{\bar{Z}}{ \nu_0  }\tilde{F} (\bar{Q})\,,
  \label{eq:rescaleprop}
\end{equation}
and the corresponding rescaled single-scale propagators are defined
via
\begin{equation}
  \Lambda \dot{G}(K)=-\frac{Z}{\Omega_{\Lambda} } \dot{\tilde{G}} (Q)\,,
  \; \; \;
  \Lambda \dot{F}(\bar{K})=-\frac{\bar{Z}}{ \nu_0  }\dot{\tilde{F}} (\bar{Q})\,.
  \label{eq:rescalesingleprop}
\end{equation}
From Eqs.~(\ref{eq:flowGammarescale1}) and
(\ref{eq:flowGammarescale2}) we can read off the scaling dimensions of
the vertices: the scaling dimension of $\tilde{\Gamma}^{(2n,m)}$ in
$D$ dimensions is
\begin{eqnarray}
  D^{(2n,m)}  =~~~~~~~~~~~~~~~~~~~~~~~~~~~~~~~~~~~~~~~~~~~~~~~~~~&& \nonumber\\[1mm]
  \left\{
    \begin{array}{ll}
      (1-n) D + z_{\text{min}} -
      ( D + z_{\varphi} ) m/2 & \mbox{for $ n \geq 1$} \\
      ( D + z_{\varphi} ) ( 1 - m/2) &\mbox{for $ n =0$}
    \end{array}
  \right.
  \,.&&\nonumber\\
  &&\label{eq:scaledim}
\end{eqnarray}
In the particular case of the Tomonaga-Luttinger model, where $D=1$
and $z_{\psi} = z_{\varphi} = 1$, we have $D^{(2n,m)} = 2 -n - m$.
Hence, in this case $\tilde{\Gamma}^{(2,0)} ( Q =0)$ and
$\tilde{\Gamma}^{(0,1)} $ are relevant with scaling dimension $+1$,
while $\tilde{\Gamma}^{ (4,0 )} (Q_i =0)$ and $\tilde{\Gamma}^{(2,1)}
( Q_i= \bar{Q}_i =0)$ are marginal. All other vertices are irrelevant.
Of course, the linear terms in the expansion of
$\tilde{\Gamma}^{(2,0)} ( Q;Q )$ for small $Q$ are also marginal.
These terms determine the wave-function renormalization factor $Z$ and
the Fermi velocity renormalization $\tilde{v}_l$, see
Eqs.~(\ref{eq:Zdefexplicit}) and (\ref{eq:vtildedef}) below.  Note
that for short-range interactions the dispersion of the zero-sound
mode is linear in any dimension.\cite{Kopietz97} Hence, as long as the
density response is dominated by the zero sound mode,
Eq.~(\ref{eq:scaledim}) remains valid for $D > 1$ with $z_{\psi} =
z_{\psi} = 1$.  In this case the scaling dimension of the purely
fermionic four-point vertex is $D^{(4,0)} = 1 - D$ and the scaling
dimension of the three-legged vertex with two fermion legs and one
boson leg is $D^{(2,1)} = (1 - D )/2$.  Both vertices become
irrelevant in $D > 1$.  As discussed in the following section, this
means that the random-phase approximation (RPA) for the effective
interaction, as well as the so-called GW approximation~\cite{Hedin65}
for the fermionic self-energy, are qualitatively correct in $D>1$.

\subsection{A simple truncation scheme: Keeping only the skeleton elements for
  two-point functions}
\label{subsec:truncation}

\begin{figure}
  \begin{center}
   \epsfig{file=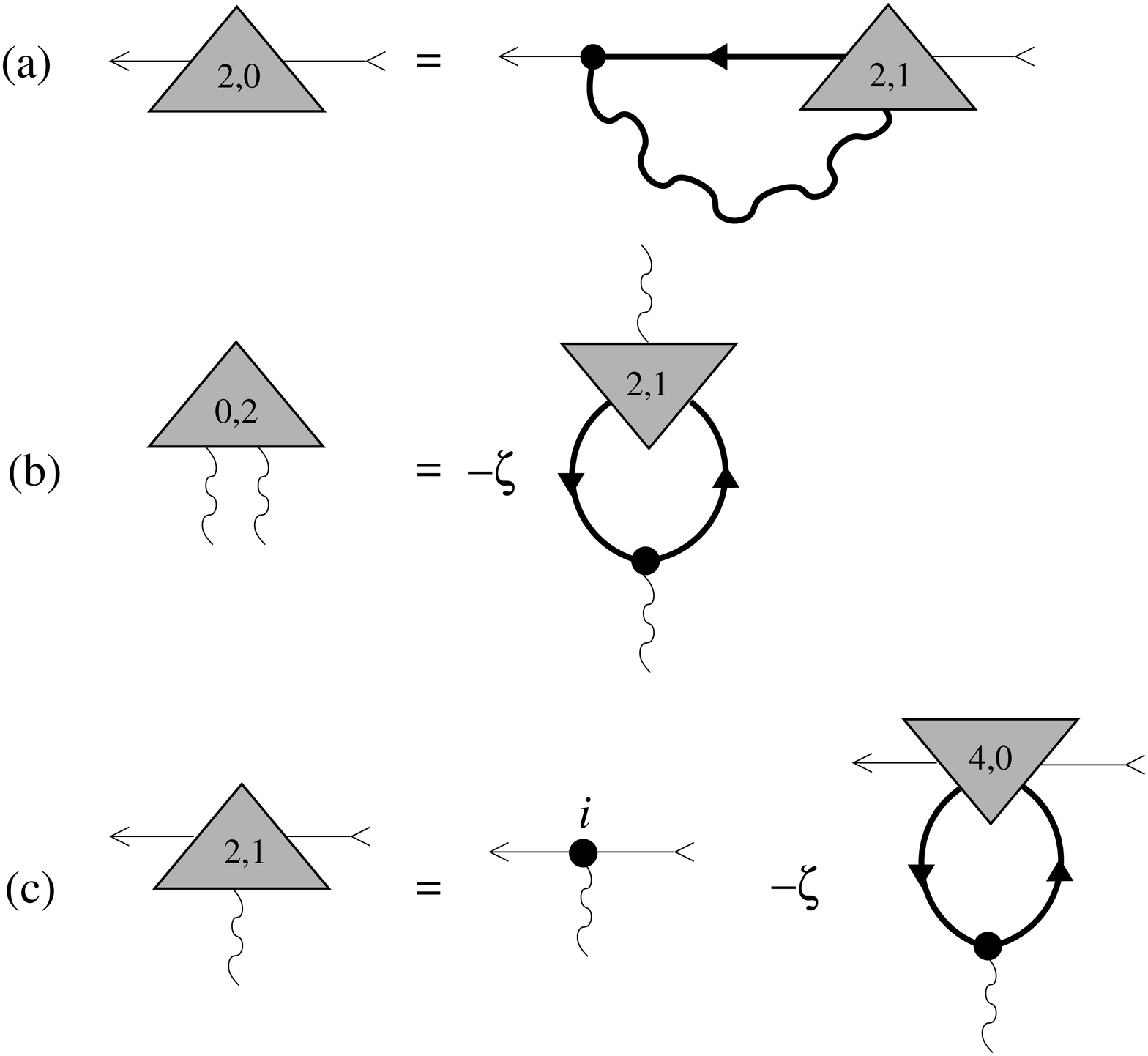,width=1.0\hsize}
  \end{center}
  \caption{Skeleton diagrams for (a) the 
    one-particle irreducible fermionic self-energy; (b) the
    one-interaction-line irreducible polarization; and (c) the
    three-legged vertex with two fermion legs and one boson leg.
    The small black circle denotes the bare three-legged vertex. Thin
    lines denote external legs.  The other graphical elements are the
    same as in Fig.~\ref{fig:picdic}.  }
  \label{fig:skeletonsigmapi}
\end{figure}

In order to solve the flow equations explicitly, one is forced to
truncate the infinite hierarchy of flow equations.  In the
one-particle irreducible version of the purely fermionic functional RG
it is common
practice\cite{Zanchi96,Halboth00,Honerkamp01,Honerkamp01b,Kampf03,Katanin04}
to retain only vertices up to the four-point vertex and set all higher
order vertices equal to zero.  Our approach offers new possibilities
for truncation schemes.  Consider the skeleton graphs
\cite{Nozieres64} for the one-particle irreducible fermionic
self-energy and the one-interaction-line irreducible polarization
shown in Figs.~\ref{fig:skeletonsigmapi}(a) and (b).
The skeleton graphs contain three basic elements: the exact fermionic
Green's function, the exact bosonic Green's function (i.e., the effective
screened interaction), and the three-legged vertex with two fermion
legs and one boson leg.  A systematic derivation of the skeleton
expansion for the vertices in our coupled Fermi-Bose theory is
presented in Appendix~B.  One advantage of our RG approach (as
compared with more conventional methods involving only fermionic
fields) is that it yields directly the flow equations for basic
elements appearing in the skeleton graphs for the self-energy and the
polarization shown in Fig.~\ref{fig:skeletonsigmapi}.  Of course, in
principle the three-legged vertex can be obtained from the vertex with
four fermion legs with the help of the skeleton graph shown in
Fig.~\ref{fig:skeletonsigmapi}(c).  However, calculating the
three-legged vertex from the four-legged vertex in this way involves
an intermediate integration, which requires the knowledge of the
momentum and frequency dependence of the four-legged vertex.
Unfortunately, in practice the purely fermionic functional RG
equations have to be severely truncated so that up to now it has not
been possible to keep track of the frequency dependence of the
four-legged fermion vertex within the purely fermionic functional RG.

\begin{figure}
  \begin{center}
    \epsfig{file=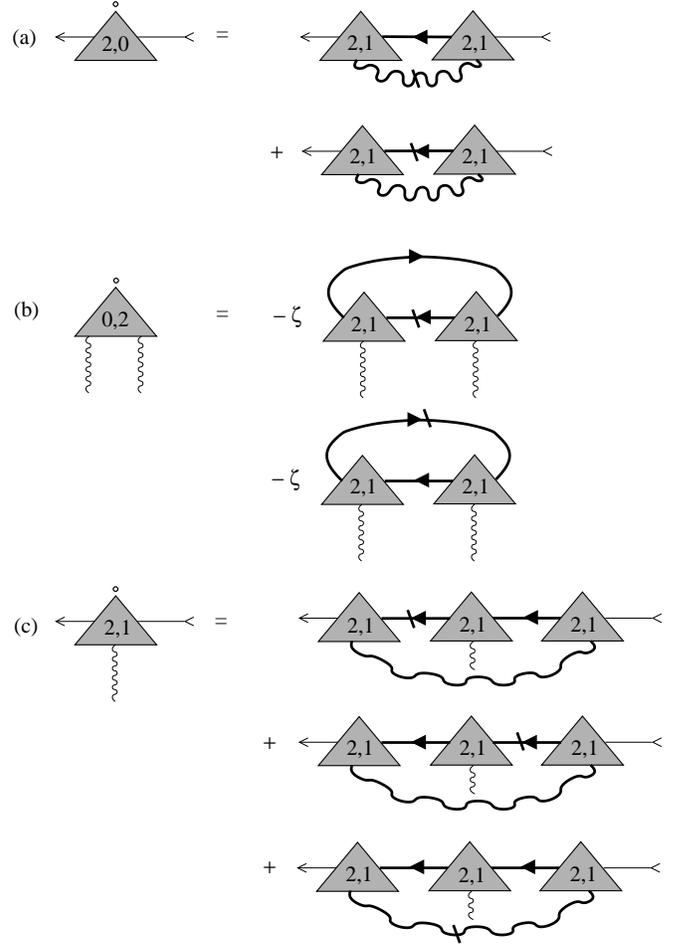,width=1.0\hsize}
  \end{center}
  \caption{Truncation of the flow equations for
    (a) fermionic self-energy, (b) irreducible polarization, and (c)
    three-legged vertex which sets all other vertices equal to zero.
    The internal lines are full propagators, which depend on the
    self-energies $\Gamma^{(2,0)} = \Sigma $ and $\Gamma^{(0,2)} =
    \Pi$.  }
  \label{fig:flowtrunc}
\end{figure}

To obtain a closed system of RG equations involving only the skeleton
elements, let us retain only the vertices $\Sigma_{\sigma} ( K )$,
$\Pi_{\sigma} ( {\bar{K}} )$ and $\Gamma^{(2,1)} ( K + \bar{K} \sigma;
K \sigma ; \bar{K} \sigma)$ on the right-hand sides of the exact flow
equations for these quantities shown in Figs.~\ref{fig:flowSigma},
\ref{fig:flowPi}, and \ref{fig:flowVCorr}, and set all other vertices
equal to zero.  The resulting closed system of flow equations is shown
graphically in Fig.~\ref{fig:flowtrunc}.
Explicitly, the flow equations are
\begin{eqnarray}
  \partial_{\Lambda} {\Sigma}_{\sigma} ( K ) & = &  
  \nonumber
  \\
  & & \hspace{-20mm}
  \int_{\bar{K}} 
  \left[   \dot{F}_{\sigma \sigma} ( \bar{K} ) G_{\sigma}  ( K + \bar{K} ) 
    + {F}_{\sigma \sigma} ( \bar{K} ) \dot{G}_{\sigma}  ( K + \bar{K} ) 
  \right]
  \nonumber 
  \\
  &  & \hspace{-20mm} \times
  \Gamma^{(2,1)} ( K + \bar{K} \sigma;   K \sigma;    \bar{K} \sigma ) 
  \Gamma^{(2,1)} ( K \sigma; K + \bar{K} \sigma; -  \bar{K} \sigma ) 
  \; ,
  \nonumber
  \\
  \label{eq:flowSigmatrunc}
\end{eqnarray}
\begin{eqnarray}
  \partial_{\Lambda} {\Pi}_{\sigma} ( \bar{K} ) & = & 
  \nonumber
  \\
  & & \hspace{-20mm} -\zeta \int_K 
  \left[ \dot{ G}_{\sigma} ( K  ) G_{\sigma} ( K + \bar{K} ) +    { G}_{\sigma} ( K  ) 
    \dot{G}_{\sigma} ( K + \bar{K} )  \right]
  \nonumber 
  \\
  &  & \hspace{-20mm} \times
  \Gamma^{(2,1)} ( K + \bar{K} \sigma;   K \sigma;    \bar{K} \sigma ) 
  \Gamma^{(2,1)} ( K \sigma; K + \bar{K} \sigma; -  \bar{K} \sigma ) 
  \; ,
  \nonumber
  \\
  \label{eq:flowPitrunc}
\end{eqnarray}
\begin{eqnarray}
  \partial_{\Lambda} {\Gamma}^{(2,1)} ( K + \bar{K} \sigma ; K \sigma ;
  \bar{K} \sigma  ) & = &  
  \nonumber
  \\
  & & \hspace{-40mm}
  \int_{\bar{K}^{\prime}} 
  \Bigl[ \dot{ F}_{\sigma \sigma} ( \bar{K}^{\prime} ) G_{\sigma} ( K + \bar{K}^{\prime} )  G_{\sigma} ( K +  \bar{K} + \bar{K}^{\prime} ) 
  \nonumber
  \\
  & & \hspace{-35mm} +
  { F}_{\sigma \sigma} ( \bar{K}^{\prime} ) \dot{G}_{\sigma} ( K + \bar{K}^{\prime} )  G_{\sigma} ( K +  \bar{K} + \bar{K}^{\prime} ) 
  \nonumber
  \\
  & & \hspace{-35mm} +
  { F}_{\sigma \sigma} ( \bar{K}^{\prime} ) G_{\sigma} ( K + \bar{K}^{\prime} )  \dot{G}_{\sigma} ( K +  \bar{K} + \bar{K}^{\prime} ) 
  \Bigr]
  \nonumber 
  \\
  &  & \hspace{-35mm} \times
  \Gamma^{(2,1)} ( K + \bar{K} \sigma ; K + \bar{K} + \bar{K}^{\prime} \sigma ; - \bar{K}^{\prime} \sigma )
  \nonumber 
  \\
  &  & \hspace{-35mm} \times
  \Gamma^{(2,1)} ( K + \bar{K} + \bar{K}^{\prime} \sigma ; K + \bar{K}^{\prime}  \sigma ;  \bar{K} \sigma )
  \nonumber 
  \\
  &  & \hspace{-35mm} \times
  \Gamma^{(2,1)} ( K + \bar{K}^{\prime} \sigma ; K  \sigma ;  \bar{K}^{\prime} \sigma )
  \; .
  \label{eq:flowGammatrunc}
\end{eqnarray}
These equations form a closed system of integrodifferential equations
that can in principle be solved numerically.  If the initial momentum
transfer cutoff $\Lambda_0$ is chosen larger than the maximal momentum
transferred by the bare interaction, and if the initial bandwidth
cutoff $v_0 \Lambda_0$ is larger than the bandwidth of the bare energy
dispersion, then the initial conditions are ${\Sigma}_{\sigma} ( K
)_{\Lambda_0} =0$, ${\Pi}_{\sigma} ( \bar{K} )_{\Lambda_0} =0$, and
${\Gamma}^{(2,1)} ( K + \bar{K} \sigma ; K \sigma ; \bar{K} \sigma
)_{\Lambda_0} = i$.  A numerical solution of these coupled equations
seems to be a difficult task, which we shall not attempt in this work.
Note, however, that in Sec.~\ref{subsec:rescale} we have argued that
for regular interactions in dimensions $D > 1$ the three-legged vertex
is actually irrelevant in the RG sense.  Hence, we expect that the
qualitatively correct behavior of the fermionic self-energy and of the
polarization can be obtained by ignoring the flow of the three-legged
vertex, setting $\Gamma^{(2,1)} \rightarrow i$.  If we further ignore
interaction corrections to the internal propagators in the flow
equation (\ref{eq:flowPitrunc}) for the polarization, it is easy to
see that the solution of this equation is nothing but the
noninteracting polarization. This is equivalent with the RPA for the
effective interaction. Substituting this into the flow equation
(\ref{eq:flowSigmatrunc}) for the self-energy and ignoring again
self-energy corrections to the internal Green's functions, we obtain the
non-self-consistent GW approximation\cite{Hedin65} for the fermionic
self-energy.  For regular interactions in $D > 1$ we therefore expect
that the RPA and the GW approximation are qualitatively correct.
However, for strong bare interactions quantitatively accurate results
can only be expected if the vertex corrections described by
Eq.~(\ref{eq:flowGammatrunc}) are at least approximately taken into
account.

We shall consider this problem again in Sec.~\ref{subsec:truncrel},
where we discuss truncations of an expansion based on relevance. To
lowest order, this approximation will agree with
Eqs.~(\ref{eq:flowSigmatrunc})--(\ref{eq:flowGammatrunc}) when the
dependence of the vertex $\Gamma^{(2,1)}$ on momenta and frequencies
is ignored. There, we use the resulting equations to calculate an
approximation to the electronic Green's function of the one-dimensional
Tomonaga-Luttinger model.  Amazingly, this simple truncation is
sufficient to reproduce the correct anomalous dimension known from
bosonization even for large values of the bare coupling.

\section{The momentum-transfer cutoff as flow parameter}
\label{sec:1dflow}

The truncation discussed in Sec.~\ref{subsec:truncation} violates the
Ward identities relating vertices with different numbers of external
legs (for a self-contained derivation of the Ward identities within
the framework of our functional integral approach; see
Appendix~\ref{sec:ward}).  Moreover, even if we do not truncate the
exact hierarchy of flow equations shown in Figs.~\ref{fig:flowSigma},
\ref{fig:flowPi}, and \ref{fig:flowVCorr}, the Ward identities are
violated for any finite value of the bandwidth-cutoff $v_0 \Lambda$,
because the cutoff leads to a violation of the underlying gauge
symmetry.  We can thus only expect the Ward identities to be restored
in the limit $v_0 \Lambda \rightarrow 0$. Recall that in the
Tomonaga-Luttinger model the Ward identities are valid in the strict
sense only in the presence of the Dirac sea, implying that the
ultraviolet cutoff $v_0\Lambda_0$ has been removed. The Ward
identities and the underlying asymptotic conservation laws are crucial
for the exact solubility of the TLM~\cite{Dzyaloshinskii74,Bohr81} and
its higher-dimensional
generalization.\cite{Metzner98,Kopietz97,Haldane92,Bartosch99} In
order to reproduce the exact solution of the TLM known from
bosonization within the functional RG, it is very important to have RG
flow equations which are consistent with the Ward identities, even for
finite values of the cutoff.  In this section we show that this
requirement is fulfilled if we work in our mixed Fermi-Bose RG with a
momentum-transfer cutoff $\Lambda$ only and take the limit $v_0
\Lambda \rightarrow 0$ of a vanishing bandwidth cutoff.

\subsection{Exact flow equations for momentum-transfer cutoff}

\begin{figure}
  \begin{center}
    \epsfig{file=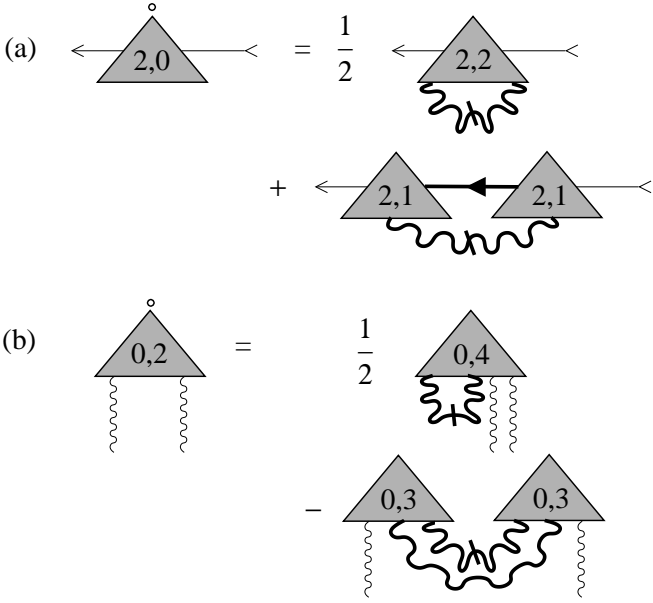,width=1.0\hsize}
  \end{center}
  \caption{
    Exact flow equations for (a) the fermionic self-energy and (b) the
    irreducible polarization in the momentum-transfer cutoff scheme.
  }
  \label{fig:flowsigmapitrans}
\end{figure}
\begin{figure}
  \begin{center}
    \epsfig{file=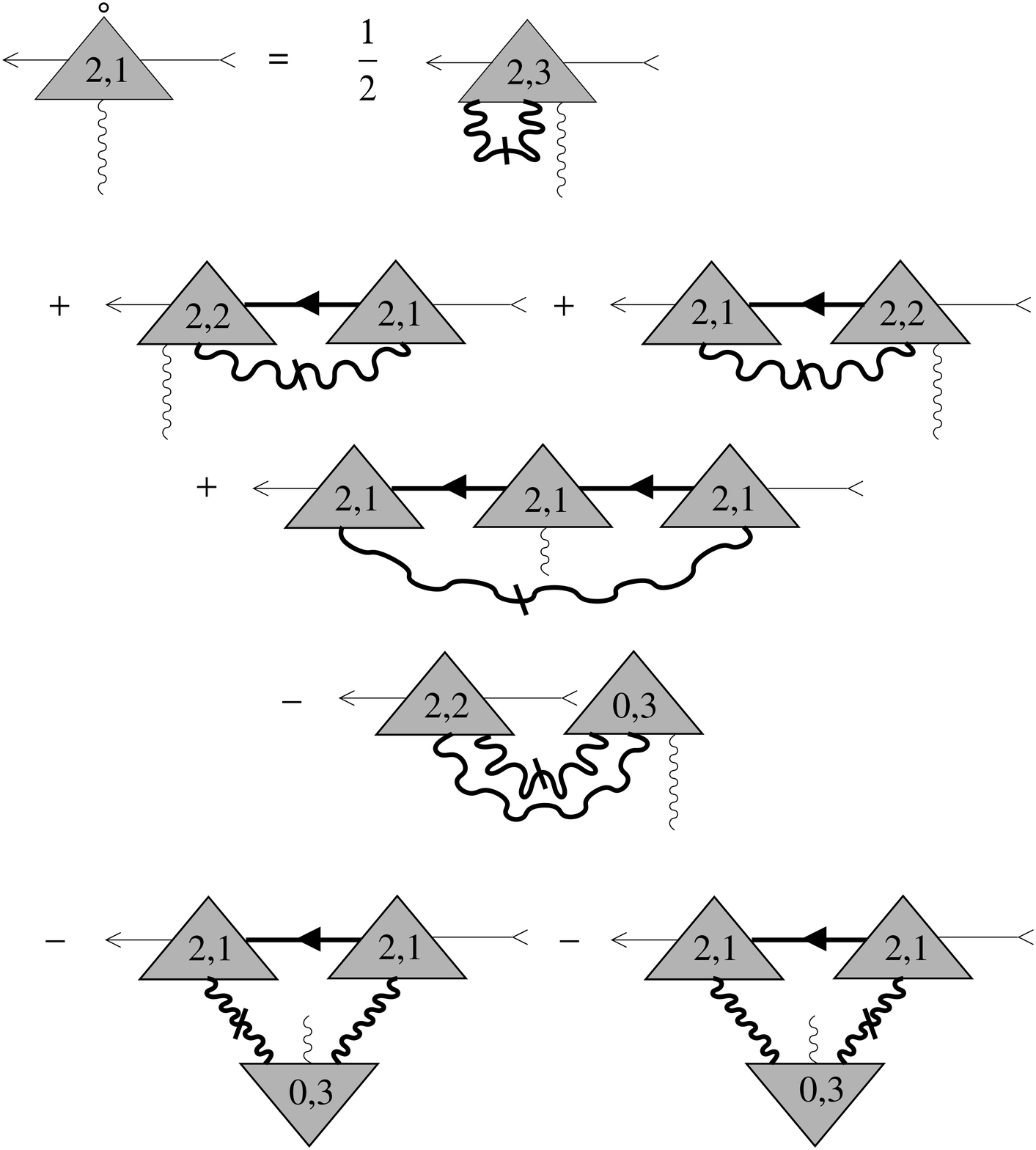,width=0.8\hsize}
  \end{center}
  \caption{
    Exact flow equations for the three-legged vertex with two
    fermion legs and one boson leg in the momentum-transfer cutoff
    scheme.  }
  \label{fig:flowGammatrans}
\end{figure}
\begin{figure}
  \begin{center}
    \epsfig{file=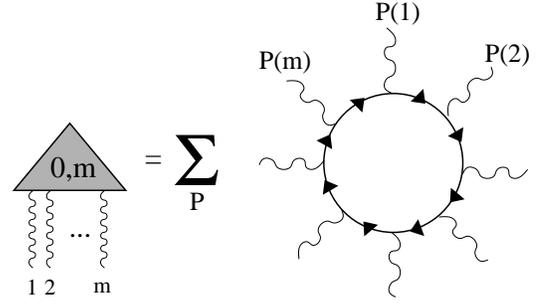,width=0.8\hsize}
  \end{center}
  \caption{
    Initial condition for the pure boson vertices in the momentum
    transfer cutoff scheme. The sum is taken over the $m!$
    permutations of the labels of the external legs. For linearized
    energy dispersion, all symmetrized closed fermion loops with more
    than two external legs vanish.  }
  \label{fig:closedloop}
\end{figure}

As already briefly mentioned at the end of
Sec.~\ref{subsec:flowphysical}, if we work with a momentum transfer
cutoff $\Lambda$ only, then all diagrams with a slash on an internal
fermionic Green's function [corresponding to the fermionic component of
the single-scale propagator given in Eq.~(\ref{eq:dotGdiagdef})] on
the right-hand sides of the exact flow equations shown in
Figs.~\ref{fig:flowSigma}, \ref{fig:flowPi}, and \ref{fig:flowVCorr}
should be omitted.  The exact flow equations for the electronic
self-energy and the irreducible polarization then reduce to
\begin{eqnarray}
  \partial_{\Lambda} \Sigma_{\sigma}(K) &=& \frac12 \int_{\bar{K}}
  \dot{F}_{\sigma\sigma}(\bar{K}) 
  \Gamma^{(2,2)}(K\sigma;K\sigma;\bar{K}\sigma,-\bar{K}\sigma)
  \nonumber\\
  && \hspace{-20mm} + \int_{\bar{K}}
  \dot{F}_{\sigma\sigma}(\bar{K})G_{\sigma}(K+\bar{K}) 
  \Gamma^{(2,1)} ( K + \bar{K} \sigma;   K \sigma;    \bar{K} \sigma ) 
  \nonumber
  \\
  &  &  \times
  \Gamma^{(2,1)} ( K \sigma; K + \bar{K} \sigma; -  \bar{K} \sigma ) 
  \; ,
  \label{eq:flowsigmatransfer}
\end{eqnarray}
\begin{eqnarray}
  \partial_{\Lambda} \Pi_{\sigma}(\bar{K}) &=& \frac12 \int_{\bar{K}'}
  \dot{F}_{\sigma\sigma}(\bar{K})
  \Gamma^{(0,4)}(\bar{K}'\sigma,-\bar{K}'\sigma,\bar{K}\sigma,-\bar{K}\sigma)
  \nonumber\\
  &&\hspace{-22mm}  
  - 
  \int_{\bar{K}'}
  \dot{F}_{\sigma\sigma}(\bar{K}')F_{\sigma\sigma}(\bar{K}+\bar{K}') 
  \Gamma^{(0,3)}(-\bar{K}\sigma,\bar{K}+\bar{K}'\sigma,-\bar{K}'\sigma)
  \nonumber\\
  &&\times
  \Gamma^{(0,3)}(\bar{K}'\sigma,-\bar{K}-\bar{K}'\sigma,\bar{K}\sigma)  \; .
  \label{eq:flowPitransfer}
\end{eqnarray}
These equations are shown graphically in
Fig.~\ref{fig:flowsigmapitrans}.  A graphical representation of the
corresponding exact flow equation for the three-legged vertex is shown
in Fig.~\ref{fig:flowGammatrans}.
Still, these flow equations look rather complicated.  Since we have
imposed a cutoff only in the momentum transfered by the bosons, the
initial conditions at scale $\Lambda_0$ are nontrivial.  The initial
value of the three-legged vertex is still $\Gamma^{(2,1)}_{\Lambda_0}
= i$, but the pure boson vertices $\Gamma^{(0,m)}$ with $m$ external
legs are initially given by the symmetrized closed fermion loops shown
in Fig.~\ref{fig:closedloop}. All other vertices vanish at the initial
scale $\Lambda_0$.
An essential simplification occurs now if we linearize the energy
dispersion relative to the Fermi surface.  If the initial momentum
transfer cutoff $\Lambda_0$ is small compared with the typical Fermi
momentum, then we may set all pure boson vertices $\Gamma^{(0,m)}$
with more than two external boson legs ($ m \geq 2$) equal to zero.
This is nothing but the closed loop
theorem,\cite{Dzyaloshinskii74,Bohr81,Kopietz97,Kopietz95,Metzner98}
which is valid exactly for the one-dimensional TLM (where the energy
dispersion is linear by definition). In higher dimensions, the closed
loop theorem is valid to a very good approximation as long as the
linearization of the energy dispersion is justified within a given
sectorization of the Fermi surface and scattering processes that
transfer momentum between different sectors of the Fermi surface can
be neglected.\cite{Kopietz97,Kopietz95} Note that the closed loop
theorem is consistent with the momentum-transfer cutoff flow, because
pure boson vertices $\Gamma^{(0,m )}$ with $ m \geq 3$ are not
generated if they initially vanish.

Assuming the validity of the closed loop theorem, the right-hand side
of the flow equation (\ref{eq:flowPitransfer}) for the polarization
vanishes identically, because it depends only on boson vertices with
more than two external legs.  Physically, this means that there are no
corrections to the noninteracting polarization, so that the RPA for
the effective interaction is exact. This is of course well known since
the pioneering work by Dzyaloshinskii and
Larkin.\cite{Dzyaloshinskii74} Moreover, the last three diagrams in
the flow equation for $\Gamma^{(2,1)}$ shown in
Fig.~\ref{fig:flowGammatrans} also vanish, because they contain the
vertex $\Gamma^{(0,3)}$.  However, the remaining diagrams in
Fig.~\ref{fig:flowsigmapitrans}(a) and Fig.~\ref{fig:flowGammatrans}
still look quite complicated, so that we still have to solve an
infinite hierarchy of coupled flow equations.  In the next subsection
we show how this infinite system of coupled integro\-differential
equations can be solved exactly.

\subsection{Ward identities as solutions of the infinite hierarchy
  of flow equations}

\begin{figure*}
  \epsfig{file=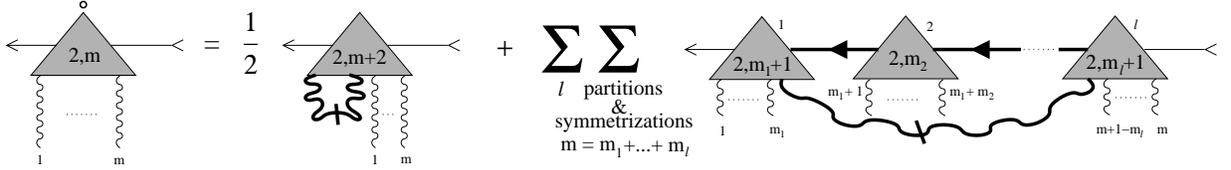,width=0.9\hsize}
  \caption{Diagrammatic representation 
    of the flow equation~(\ref{eq:flowdisord}) of vertices with two
    fermion legs and a general number of boson legs provided the pure
    boson vertices with more than two external legs vanish, as implied
    by the closed loop theorem.}
  \label{fig:flow_2Fermi_mBoson}
\end{figure*}

Let us consider the terms on the right-hand sides of the flow
equations for the vertices $\Gamma^{(2,m)}$ with two external fermion
legs and an arbitrary number of boson legs.  Assuming the validity of
the closed loop theorem, all pure boson vertices $\Gamma^{(0,m)}$ with
$m \geq 2$ vanish. From Fig.~\ref{fig:flowsigmapitrans}(a) and
Fig.~\ref{fig:flowGammatrans} it is clear that in general the
right-hand side of the flow equation for $\partial_{\Lambda}
\Gamma^{(2,m)}$ depends on $\Gamma^{(2, m+2 )}$ and on all
$\Gamma^{(2,m^{\prime})}$ with $m^{\prime} \leq m$.  In fact, from our
general expression for the flow of the totally symmetrized vertices
given in Eq.~(\ref{eq:flow_vert}), we can derive the flow equations for
the vertices $\Gamma^{(2,m)}$ with arbitrary $m$ in closed form (we
omit for simplicity the degeneracy index $\sigma$),
\begin{widetext}
  \begin{eqnarray}
    \partial_{\Lambda}\Gamma^{(2,m)}(K';K;\bar{K}_1,\dots,\bar{K}_m)
    = \frac12\int_{\bar{K}}\dot{F}_{\sigma\sigma}(\bar{K})
    \Gamma^{(2,m+2)}(K';K;-\bar{K},\bar{K},\bar{K}_1,\dots,\bar{K}_m)
    +\sum_{l=2}^{\infty}\sum_{m_1,\dots,m_l=1}^{\infty}\!\!\frac{\delta_{m,\sum_im_i}}{\prod _i m_i!}
    &&
    \nonumber\\
    \times\sum_{P}\int_{\bar{K}}
    \dot{F}(\bar{K})
    \Gamma^{(2,m_1+1)}\left(K';\tilde{K}_1;\bar{K}_{P(1)},\dots,\bar{K}_{P(m_1)},-\bar{K}\right)
    G(\tilde{K}_1)
    \Gamma^{(2,m_2)}\left(\tilde{K}_1;\tilde{K}_2;\bar{K}_{P(m_1+1)},\dots,\bar{K}_{P(m_1+m_2)}\right)
    &&
   \nonumber\\
    \times G(\tilde{K}_2)
    \cdot\ldots\cdot
    G(\tilde{K}_{l-1})
    \Gamma^{(2,m_l+1)}\left(\tilde{K}_{l-1};K;\bar{K},\bar{K}_{P(m-m_l+1)},\dots,\bar{K}_{P(m)}\right)
    \; ,~~~~~~~~~~~~~~~~~~~~~~~&&
    \label{eq:flowdisord}
  \end{eqnarray}
\end{widetext}
where we have defined
\begin{equation}
  K' = K + \sum_{i=1}^m \bar{K}_i\,, 
  \qquad
  \tilde{K}_i = K' + \bar{K} - \sum_{j=1}^{m_1+\ldots+m_i} \bar{K}_{P(j)}\,,
\end{equation}
and $P$ denotes a permutation of $\{1,\dots,m\}$.  A graphical
representation of Eq.~(\ref{eq:flowdisord}) is shown in
Fig.~\ref{fig:flow_2Fermi_mBoson}.
Note that the flow equation (\ref{eq:flowsigmatransfer}) for the
irreducible self-energy is a special case of Eq.~(\ref{eq:flowdisord})
for $m=0$.

\begin{figure}[b]
  \begin{center}
    \epsfig{file=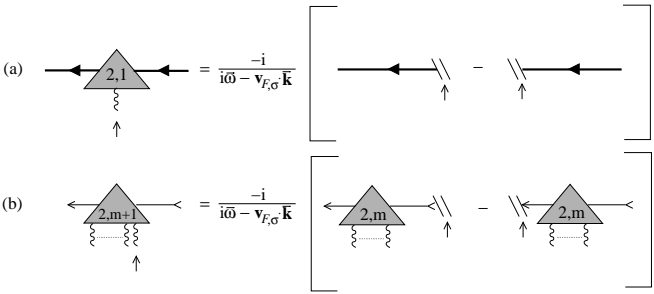,width=1.0\hsize}
  \end{center}
  \caption{
    (a) Diagrammatic representation of the Ward identity (\ref{eq:WI1})
    for the three-legged vertex and (b) of the Ward identity
    (\ref{eq:WIm}) for the vertex with two fermion legs and $m>1$
    boson legs.  The small arrow indicates the place in the diagram
    where the external bosonic energy-momentum enters.  A double slash
    to the right of an arrow means that the bosonic momentum is added
    before the corresponding Green's function, while a double slash to
    the left of an arrow means that the momentum is added after the
    Green's function.  }
  \label{fig:ward}
\end{figure}

We are now facing the problem of solving the infinite hierarchy of
coupled flow equations given by Eq.~(\ref{eq:flowdisord}).  In view of
the fact that these equations are exact and that in one dimension the
single-particle Green's function of the TLM can be calculated exactly
via bosonization, we expect that this infinite hierarchy of flow
equations can also be solved exactly.  Indeed, the solutions of these
equations are nothing but infinitely many Ward identities relating the
vertex $\Gamma^{(2,m)}$ with two fermion legs and $m$ boson legs to
the vertex $\Gamma^{(2, m-1)}$ with one boson leg less.  We derive
these Ward identities within the framework of our functional integral
approach in Appendix~C.  For $m =1$ the Ward identity is well
known~\cite{Dzyaloshinskii74,Bohr81,Metzner98,Kopietz97}
\begin{eqnarray}
  G ( K + \bar{K} ) \Gamma^{(2,1)} ( K + \bar{K} ; K ; \bar{K} )  G ( K )
  & = &
  \nonumber
  \\
  & & \hspace{-55mm} = 
  \frac{-i}{i\bar{\omega} -  {\bf{v}}_{F, \sigma} \cdot \bar{\bf{k}}}
  \Big[ G ( K + \bar{K} ) - G ( K ) \Big]
  \; .
  \label{eq:WI1}
\end{eqnarray}
Here ${\bf{v}}_{F, \sigma}$ is the Fermi velocity associated with the
independent fermionic label $K = ({\bf{k}} , i \omega )$, where $|
{\bf{k}} - {\bf{k}}_{ F , \sigma} | \ll | {\bf{k}}_{ F , \sigma} | $.
The Ward identity (\ref{eq:WI1}) has been used in
Refs.~\onlinecite{Dzyaloshinskii74} and \onlinecite{Metzner98} to
close the skeleton equation for the self-energy and thus obtain the
exact Green's function of the TLM without invoking the machinery of
bosonization. A Ward identity for $\Gamma^{(4,1)}$ has also been used
to prove the vanishing of the renormalization group $\beta$ function
for the TLM.\cite{DiCastro91} However, for solving the TLM exactly
within the framework of the functional RG, we need the Ward identities
for all vertices $\Gamma^{(2,m)}$ with $m \geq 1$.  As shown in
Appendix C, for linear energy dispersion we have
\begin{widetext}
  \begin{eqnarray}
    \Gamma^{(2,m)} \Big(K';K;\bar{K}_1,\dots,\bar{K}_m\Big)
    =
    \frac{-i}{i\bar{\omega}_l -  {\bf{v}}_{F,\sigma} \cdot \bar{\bf{k}}_l}
    \Bigg[
    \Gamma^{(2,m-1)} \Big(K';K+\bar{K}_l;
    \bar{K}_1,\dots,\bar{K}_{l-1},\bar{K}_{l+1},\dots,\bar{K}_m\Big)
    &&\nonumber\\
    -
    \Gamma^{(2,m-1)} \Big(K'-\bar{K}_l;K;
    \bar{K}_1,\dots,\bar{K}_{l-1},\bar{K}_{l+1},\dots,\bar{K}_m\Big)
    \Bigg] \; , &&
    \label{eq:WIm}
  \end{eqnarray}
  where $1 \leq l \leq m$.  For clarity let us write down here the
  special case $m=2$,
  \begin{eqnarray}
    \Gamma^{(2,2)} \Big(K + \bar{K}_1 + \bar{K}_2 ;K;\bar{K}_1,\bar{K}_2\Big)
    =
    \frac{-i}{i\bar{\omega}_1 -  {\bf{v}}_{F , \sigma} \cdot \bar{\bf{k}}_1}
    \Big[
    \Gamma^{(2,1)} \Big(K  + \bar{K}_1 + \bar{K}_2 ; K+\bar{K}_1;
    \bar{K}_2\Big)
    -
    \Gamma^{(2,1)} \Big(K + \bar{K}_2;K;
    \bar{K}_2\Big)
    \Big] 
    &&
    \nonumber
    \\
    &&  \hspace{-140mm}  =   \frac{-i}{i\bar{\omega}_2 -  
      {\bf{v}}_{F, \sigma} \cdot \bar{\bf{k}}_2}
    \Big[
    \Gamma^{(2,1)} \Big(K  + \bar{K}_1 + \bar{K}_2 ; K+\bar{K}_2;
    \bar{K}_1\Big)
    -
    \Gamma^{(2,1)} \Big(K + \bar{K}_1;K;
    \bar{K}_1\Big)
    \Big] 
    \; . 
    \label{eq:WI2}
  \end{eqnarray}
\end{widetext}
Diagrammatic representations of the Ward identities given in
Eqs.~(\ref{eq:WI1}) and (\ref{eq:WIm}) are shown in
Fig.~\ref{fig:ward}.
To prove that these Ward identities indeed solve our infinite system
of flow equations given by Eqs.~(\ref{eq:flowsigmatransfer}) and
(\ref{eq:flowdisord}), we start from the flow equation for
$\Gamma^{(2, m+1)}$.  Substituting on both sides of this exact flow
equation the Ward identities, we can reduce it to a new flow equation
involving only vertices where the number of boson legs is reduced by
one, but with an external bosonic momentum entering the vertices at
various places.  Graphically, we indicate the place where the bosonic
momentum enters the vertex by a double slash, as shown in
Fig.~\ref{fig:ward}.  The important point is now that all diagrams
with double slashes attached to intermediate Green's functions cancel
due to the fact that all vertices $\Gamma^{(2,m)}$ can be expressed in
terms of a difference of vertices $\Gamma^{(2,m-1)}$, with a same
prefactor that is independent of $m$. Graphically, only the diagrams
with a double slash attached to the leftmost or rightmost Green's
function survive.  Canceling the common prefactor, it is then easy to
see that the RG equation derived in this way from the functional RG
equation for $\Gamma^{(2,m+1)}$ is nothing but the exact RG equation
for $\Gamma^{(2,m)}$.  Hence, the Ward identities provide relations
between the vertices $\Gamma^{(2,m)}$ that are consistent with the
relations implied by the exact hierarchy of RG flow equations in the
momentum-transfer cutoff scheme.  In other words, the Ward identities
are the solutions of the infinite hierarchy of flow equations!

\subsection{Exact solution of the Tomonaga-Luttinger model via the exact 
  RG}
\label{subsec:exactTLM}

Given the cascade of Ward identities (\ref{eq:WI1}) and (\ref{eq:WIm})
we can close the integro\-differential equation
(\ref{eq:flowsigmatransfer}) for the irreducible self-energy.  Note
that this equation involves both the three-legged vertex and the
four-legged vertex with two fermion legs and two boson legs, so that
the Ward identity (\ref{eq:WI1}) is not sufficient to close the flow
equation.  Of course, if one is only interested in calculating the
Green's function of the TLM, it is simpler to start from the skeleton
equation for the self-energy shown in Fig.~\ref{fig:skeletonsigmapi},
which can be closed by means of the Ward identity (\ref{eq:WI1}) for
the three-legged vertex only. Nevertheless, it is instructive to see
how the exact solution emerges within the framework of the functional
RG.  Substituting Eqs.~(\ref{eq:WI1}) and (\ref{eq:WI2}) into
Eq.~(\ref{eq:flowsigmatransfer}), we obtain the following
integro\-differential equation for the electronic self-energy:
\begin{eqnarray}
  \partial_{\Lambda}\Sigma_{\sigma} (K) & = & G_{\sigma}^{-2}(K)
  \int_{\bar{K}}\frac{\dot{F}_{\sigma \sigma}(\bar{K})}{
    (i\bar{\omega}- {\bf{v}}_{F, \sigma} \cdot \bar{\bf{k}} )^2}
  \nonumber
  \\
  & & \times
  \left[G_{\sigma}(K)-G_{\sigma}(K+\bar{K})\right]\,.
  \label{eq:sigmaflowTLM}
\end{eqnarray}
Here the index $\sigma$ labels not only the different spin species,
but also the different patches of the sectorized Fermi
surface.\cite{Kopietz97} For example, for the spinless case $\sigma =
\pm k_F$.  Using the fact that in the momentum-transfer cutoff scheme
$G^2 \partial_{\Lambda} \Sigma = \partial_{\Lambda} G$ we can
alternatively write Eq.~(\ref{eq:sigmaflowTLM}) as a {\it{linear}}
integro\-differential equation for the fermionic Green's function,
\begin{eqnarray}
  \partial_{\Lambda}G_{\sigma}(K) 
  & = &
  \int_{\bar{K}}\frac{\dot{F}_{\sigma\sigma}(\bar{K})}{(i\bar{\omega}-
    {\bf{v}}_{F ,\sigma} \cdot \bar{\bf{k}} )^2}   
  \nonumber
  \\
  & & \times
  \left[G_{\sigma}(K)-G_{\sigma}(K+\bar{K})\right]\,.
  \label{eq:Gflow_ex}
\end{eqnarray}
If we had simply set the vertex $\Gamma^{(2,2)}$ equal to zero in
Eq~(\ref{eq:flowsigmatransfer}) and had then closed this equation by
means of the Ward identity (\ref{eq:WI1}), we would have obtained a
nonlinear equation. Thus, the linearity of Eq.~(\ref{eq:Gflow_ex}) is
the result of a cancellation of nonlinear terms arising from both
Ward identities (\ref{eq:WI1}) and (\ref{eq:WI2}).  Because the second
term on the right-hand side of Eq.~(\ref{eq:Gflow_ex}) is a
convolution, we can easily solve this equation by means of a Fourier
transformation to imaginary time and real space.  Defining
\begin{eqnarray}   
  G_{\sigma} ( X ) & = & 
  \int_K e^{ i ( {\bf{k}} \cdot {\bf{r}} - \omega \tau )} 
  G_{\sigma }(K)\,,
  \label{eq:FTGdef}
  \\
  H_{ \Lambda , \sigma } ( X )
  & = &
  \int_{\bar{K}} e^{ i ( \bar{\bf{k}} \cdot {\bf{r}} - \bar{\omega} \tau )} 
  \frac{\dot{F}_{\sigma\sigma}(\bar{K})}{(i\bar{\omega}-
    {\bf{v}}_{F ,\sigma} \cdot \bar{\bf{k}} )^2}   
  \; ,
  \label{eq:Hsigmadef}
\end{eqnarray}
where $ X = (\tau , {\bf{r}} )$, the flow equation (\ref{eq:Gflow_ex})
is transformed to
\begin{equation}
  \big[\partial_{\Lambda}+H_{ \Lambda , \sigma }(X) - H_{ \Lambda , \sigma }(0)\Big]G_{\sigma}(X) = 0\,.
\end{equation}
This implies the conservation law
\begin{equation}
  \partial_{\Lambda}\left[
    \exp\left\{
      \int_0^{\Lambda}d\Lambda'\left[
        H_{ \Lambda' , \sigma }(X)-H_{ \Lambda' , \sigma }(0)
      \right]
    \right\}
    G_{\sigma}(X)
  \right] = 0\,.
\end{equation}
Integrating from $\Lambda=0$ to $\Lambda=\Lambda_0$, we obtain
\begin{equation}
  G_{\sigma}(X)=G_{0 , \sigma}(X)\,\exp\left[Q_{\sigma}(X)\right]\,,
  \label{eq:ex_sol}
\end{equation}
with
\begin{equation}
  Q_{\sigma}(X) = S_{\sigma}(0) - S_{\sigma}(X)\,,
  \label{eq:Qdef}
\end{equation}
and
\begin{eqnarray}
  S_{\sigma}(X)& = & - \int_0^{\Lambda_0}d\Lambda'H_{\Lambda', \sigma }(X)
  \nonumber
  \\
  &  & \hspace{-15mm} = \int_{\bar{K}}\frac{\Theta(\Lambda_0 - |\bar{\bf{k}}| )
    F_{\sigma\sigma}(\bar{K})}
  {(i\bar{\omega}-   {\bf{v}}_{F ,\sigma}   \cdot \bar{\bf{k}} )^2} 
  \cos(    \bar{\bf{k}} \cdot {\bf{r}} -  \bar{\omega} \tau)\,,
  \label{eq:DW}
\end{eqnarray}
where we have used the invariance of the RPA interaction $F ( \bar{K}
)$ under $\bar{K}\to-\bar{K}$. The solution in
Eqs.~(\ref{eq:ex_sol})--(\ref{eq:DW}) is well known from the
functional integral approach to bosonization
\cite{Kopietz97,Kopietz95,Fogedby76,Lee88} where $Q_{\sigma}(X)$
arises as a Debye-Waller factor from Gaussian averaging over the
distribution of the Hubbard-Stratonovich field.  In one dimension,
Eqs.~(\ref{eq:ex_sol})--(\ref{eq:DW}) can be shown \cite{Kopietz97} to
be equivalent to the exact solution for the Green's function of the
Tomonaga-Luttinger model obtained via conventional bosonization.  

Once the exact single-particle Green's function is known, the Ward
identities in Eqs.~(\ref{eq:WI1}) and (\ref{eq:WIm}) iteratively yield
expressions for the vertices $\Gamma^{(2,m)}$ that solve the whole
hierarchy of flow equations for the vertices with two fermion and an
arbitrary number of boson legs. In principle, the method described in
this section can be applied also to vertices with more than two
fermion legs. For example, the right-hand sides of the flow equations for the
vertices $\Gamma^{(4,m)}$ contain only vertices with no more than four
fermion legs. Ward identities for these vertices would again yield a
solution of this complete hierarchy, once the vertices
$\Gamma^{(2,m)}$ are known. This procedure can be iterated to obtain
vertices with an arbitrary number of external legs using at each step
the complete flow of vertices with two fewer fermion legs obtained in
the previous step.  We have thus devised a method to obtain all
correlation functions of the TLM entirely within the framework of the
functional RG.

\subsection{Truncation scheme based on relevance}
\label{subsec:truncrel}

The structure of the exact Green's function of the TLM and the
corresponding spectral function $A ( k , \omega )=-\pi^{-1}\text{Im}\,
G(k,\omega+i0)$ depend crucially on the Ward identities
discussed above, which in turn are only valid if the energy dispersion
is strictly linear.  In order to assess the validity of the
linearization of the energy dispersion, it is important to develop
truncations of the exact hierarchy of flow equations that do not
explicitly make use of the validity of the asymptotic Ward identities.
We now propose such a truncation scheme.

The coefficients generated in the expansion of a given vertex
$\Gamma^{(2n,m)}(K'_1,\dots,K'_n; K_1, \dots,
K_n;\bar{K}_1,\dots,\bar{K}_m) $ in powers of frequencies and momenta
have decreasing scaling dimensions, so that the most relevant part of
any vertex is obtained by setting all momenta and frequencies equal to
zero.  This classification leads to a simple truncation scheme: We
retain only those vertices whose leading (momentum- and
frequency-independent) part has a positive or vanishing scaling
dimension, corresponding to relevant or marginal couplings in the
usual RG jargon.  In the context of calculating the critical
temperature of the weakly interacting Bose gas in three dimensions,
such a truncation procedure has recently been shown to give very
accurate results.\cite{Ledowski04}

To begin with, let us classify all couplings according to their
relevance.  With the rescaling defined in Sec.~\ref{subsec:rescale},
for $D = z_{\psi} = z_{\varphi} = 1$, the scaling dimensions of the
vertices $\tilde{\Gamma}^{(2n,m)}$ are $D^{(2n,m)} = 2 -n -m$, see
Eq.~(\ref{eq:scaledim}). Hence the vertex $\tilde{\Gamma}^{(2,2)}$ as
well as the vertices $\tilde{\Gamma}^{(0,3)}$ and
$\tilde\Gamma^{(0,4)}$, whose unrescaled versions appear on the
right-hand sides of Eqs.~(\ref{eq:flowsigmatransfer}) and
(\ref{eq:flowPitransfer}), are irrelevant in the RG sense. In
contrast, the momentum- and frequency-independent part of the
three-legged vertex,
\begin{equation}
  \tilde{\gamma}_l = \tilde{\Gamma}^{(2,1)} ( 0; 0;0) = \left(  \frac{\Lambda}{\nu_0 \Omega_{\Lambda} } 
  \right)^{1/2}  Z_l \Gamma^{(2,1)} ( K_F ; K_F ; 0)
  \; ,
  \label{eq:tildegdef}
\end{equation}
is marginal.\cite{Ueda84}  Here $K_F = ( \pm k_F , \omega =0)$.  From
the general flow equations (\ref{eq:flowGammarescale1}) and
(\ref{eq:inhrescale1}) for the rescaled vertices we obtain the
following exact flow equation for the rescaled self-energy defined in
Eq.~(\ref{eq:Sigmarescaledef}):
\begin{equation}
  \partial_l \tilde{\Sigma}_l ( Q ) =  \left( 1 - \eta_l + Q \cdot \frac{ \partial}{\partial Q} \right) 
  \tilde{\Sigma}_l ( Q ) + \dot{\tilde{\Gamma}}^{(2,0)}_l ( Q )
  \label{eq:flowsigma}
  \; ,
\end{equation}
with [see Eq.~(\ref{eq:inhrescale1})]
\begin{equation}
  \dot{\tilde{\Gamma}}^{(2,0)}_l ( Q ) = - \frac{Z_l}{\Omega_{\Lambda}} \Lambda \partial_{\Lambda}
  \Gamma_{\Lambda}^{(2,0)} ( K )
  \; .
\end{equation}
We restrict ourselves to spinless fermions here and choose
$\Omega_{\Lambda} = \bar{\Omega}_{\Lambda} = v_F \Lambda$, so that
with $\nu_0=(\pi v_F)^{-1}$ the prefactor in Eq.~(\ref{eq:tildegdef})
turns out to be $ ( \frac{\Lambda}{\nu_0 \Omega_{\Lambda} } )^{1/2} =
\pi^{1/2}$.  As usual, the fermionic wave-function renormalization
factor $Z_l$ is defined via
\begin{equation}
  Z_l = \left[ 1 - \left. \frac{\partial \Sigma ( K ) }{\partial ( i \omega ) } \right|_{ K =0} \right]^{-1}
  = 1 + \left.  \frac{\partial \tilde{\Sigma}_l ( Q ) }{\partial ( i \epsilon ) } \right|_{ Q =0}
  \label{eq:Zdefexplicit}
  \; .
\end{equation}
According to Eq.~(\ref{eq:etadef}) the wave-function renormalization
$Z_l$ satisfies the flow equation
\begin{equation}
  \partial_lZ_l= -\eta_l Z_l\;,
\end{equation}
where the flowing anomalous dimension of the fermion fields is given
by
\begin{equation}
  \eta_l = - \left. \frac{\partial \dot{\tilde{\Gamma}}^{(2,0)}_l ( Q ) }{\partial ( i \epsilon ) } \right|_{ Q =0}
  \; .
  \label{eq:etaexplicit}
\end{equation}
According to Eq.~(\ref{eq:flowsigma}) the constant part of the
self-energy,
\begin{equation}
  \tilde{r}_l = \tilde{\Sigma}_l ( 0 )
  \; ,
  \label{eq:rldef}
\end{equation}
is relevant and satisfies
\begin{equation}
  \partial_l \tilde{r}_l =  \left( 1 - \eta_l  \right) 
 \tilde{r}_l + \dot{\tilde{\Gamma}}^{(2,0)}_l ( 0 )
 \label{eq:flowr}
 \; .
\end{equation}
In general, $\tilde{r}_l$ will only flow into the fixed point if the
initial coupling $\tilde{r}_0$ is properly fine-tuned.  Apart from
$Z_l$, there are two more marginal couplings.  The first is the Fermi
velocity renormalization factor \cite{Busche02}
\begin{equation}
  \tilde{v}_l = Z_l + \left.
    \frac{\partial \tilde{\Sigma}_l ( Q ) }{\partial  q } \right|_{ Q =0}
  \; ,
  \label{eq:vtildedef}
\end{equation}
and the second marginal coupling is the momentum- and
frequency-independent part $\tilde{\gamma}_l$ of the rescaled
three-legged vertex given in Eq.~(\ref{eq:tildegdef}).  The exact flow
equations for $\tilde{v}_l$ and $\tilde{\gamma}_l$ are
\begin{equation}
  \partial_l \tilde{v}_l =   - \eta_l  \tilde{v}_l 
  + \left. \frac{ \partial \dot{\tilde{\Gamma}}^{(2,0)}_l ( Q )}{\partial q }
  \right|_{ Q=0}
  \label{eq:flowv}
  \; ,
\end{equation}
and
\begin{equation}
  \partial_l \tilde{\gamma}_l =   - \eta_l  \tilde{\gamma}_l 
  + \dot{\tilde{\Gamma}}^{(2,1)}_l ( 0 ; 0;0)
  \label{eq:flowg}
  \; .
\end{equation}
If we retain only relevant and marginal couplings, then in the
momentum-transfer cutoff scheme the rescaled fermionic Green's function
defined in Eq.~(\ref{eq:rescaleprop}) is in $D=1$ simply approximated
by
\begin{equation}
  \tilde{G} ( Q ) \approx \frac{1}{i  \epsilon - \tilde{v}_l q - \tilde{r}_l }
  \; .
  \label{eq:Gtildeapprox}
\end{equation}
In order to make progress, we have to approximate the inhomogeneities
$\dot{\tilde{\Gamma}}^{(2,0)}_l ( Q )$ and
$\dot{\tilde{\Gamma}}^{(2,1)}_l ( 0 ; 0;0)$.  In
Sec.~\ref{subsec:truncation} we have proposed an approximation scheme
which retains only the skeleton elements of the two-point functions.
In the momentum-transfer cutoff scheme, the corresponding flow
equations (\ref{eq:flowSigmatrunc},\ref{eq:flowPitrunc},
\ref{eq:flowGammatrunc}) further simplify because we should omit all
terms involving the fermionic single-scale propagator.  Unfortunately,
the resulting nonlinear integro\-differential equations still cannot
be solved analytically.  In order to simplify these equations further,
let us replace the three-legged vertex on the right-hand sides of
these equations by its marginal part.  In this approximation we obtain,
from Eq.~(\ref{eq:flowSigmatrunc}),
\begin{equation}
  \dot{\tilde{\Gamma}}^{(2,0)}_l ( Q ) \approx 
  \tilde{\gamma}_l^2  \int_{ \bar{Q}} 
  \dot{ \tilde{F}} ( \bar{Q} ) \tilde{G} ( Q + \bar{Q} )
  \label{eq:dotGtrunc}
  \; ,
\end{equation}
and from Eq.~(\ref{eq:flowGammatrunc}),
\begin{equation}
  \dot{\tilde{\Gamma}}^{(2,1)}_l ( 0;0;0 ) \approx 
  \tilde{\gamma}_l^3  \int_{ \bar{Q}} 
  \dot{ \tilde{F}} ( \bar{Q} ) \tilde{G}^2 ( \bar{Q} )
  \label{eq:dotGammatrunc}
  \; .
\end{equation}
In order to be consistent, we should approximate $\tilde{G} ( Q )$ in
Eqs.~(\ref{eq:dotGtrunc}) and (\ref{eq:dotGammatrunc}) by
Eq.~(\ref{eq:Gtildeapprox}).  Then it is easy to see that the second
term on the right-hand sides of the flow equations (\ref{eq:flowv})
and (\ref{eq:flowg}) exactly cancels the contribution from the
anomalous dimension, so that
\begin{equation}
  \partial_l \tilde{\gamma}_l = 0
  \;, \; \; \; \partial_l \tilde{v}_l = 0
  \; .
  \label{eq:flowgzero}
\end{equation}
For explicit calculations, let us assume that the usual couplings of
the TLM\cite{Solyom79} are $g_2 = g_4 = f_0$, so that
\begin{equation}
  \dot{ \tilde{F}} ( \bar{Q} ) = 
  \delta ( 1 - | \bar{q} | )
  \frac{ \tilde{f}_0 (\bar{q}^2 + \bar{\epsilon}^2)}{ ( 1 + \tilde{f}_0) \bar{q}^2 + \bar{\epsilon}^2}
  \; ,
  \label{eq:frpatlm}
\end{equation}
where $\tilde{f}_0 = \nu_0 f_0$.  From Eqs.~(\ref{eq:etaexplicit}) and
(\ref{eq:dotGtrunc}) we then find that the anomalous dimension
$\eta=\eta_l$ does not flow and is given by
\cite{footnoteeta}
\begin{equation}
  \eta = \frac{ \tilde{f}_0^2}{ 2 \sqrt{ 1 + \tilde{f}_0} 
    \left[ \sqrt{ 1 + \tilde{f}_0} + 1 \right]^2}
  \; ,
  \label{eq:etares}
\end{equation}
which agrees exactly with the bosonization
result.\cite{Kopietz97,footnoteVolker} We emphasize that
Eq.~(\ref{eq:etares}) is the correct anomalous dimension of the TLM,
even for $\tilde{f}_0 \gg 1$, so that, at least as far as the
calculation of $\eta$ is concerned, the validity of our simple
truncation is not restricted to the weak coupling regime.  Recall that
the restriction to weak coupling is one of the shortcomings of the
conventional fermionic functional
RG,\cite{Zanchi96,Halboth00,Honerkamp01,Honerkamp01b,Kopietz01,Busche02,Ledowski03,Tsai01,Binz02,Meden02,Kampf03,Katanin04}
which was implemented for the TLM in Ref.~\onlinecite{Busche02}.
Because $\eta$ is finite, the running vertex $ \Gamma^{(2,1)} ( K_F ;
K_F ; 0)$ without wave-function renormalization actually diverges for
$\Lambda \rightarrow 0$. However, the properly renormalized vertex
$\tilde{\gamma}_l \propto Z_l \Gamma^{(2,1)} ( K_F ; K_F ; 0 )$
remains finite due to the vanishing wave-function renormalization
\begin{equation}
  Z_l=e^{-\eta l}=\left(\frac{\Lambda}{\Lambda_0}\right)^{\eta}
  \;
\end{equation}
for $l\to\infty$.  Integrating the flow equation (\ref{eq:flowsigma})
for the self-energy with the inhomogeneity approximated by
Eqs.~(\ref{eq:dotGtrunc}) and (\ref{eq:Gtildeapprox}), we obtain,
after going back to physical variables \cite{footnoteeta}
\begin{eqnarray}
  \Sigma ( k_F + k , i \omega ) = ~~~~~~~~~~~~~~~~~~~~~~~~~~~~~~~~~~~~~~~~~~~~~&& 
  \nonumber\\
  - \int_{ - \Lambda_0}^{\Lambda_0}
  \frac{ d \bar{k}}{ 2 \pi} \int_{ - \infty}^{\infty} 
  \frac{ d \bar{\omega}}{ 2 \pi} 
  \left(  \frac{\Lambda_0}{   | \bar{k} | } \right)^{\eta}
  \frac{   f^{\rm RPA} (  \bar{k} , i \bar{\omega} )}{ i ( \omega + \bar{\omega} )
    - v_F ( k + \bar{k} )   }\; ,&&
  \nonumber\\
  \label{eq:sigmatruncres} 
\end{eqnarray}
where the RPA screened interaction is
\begin{equation}
  f^{\rm RPA} (  \bar{k} , i \bar{\omega} ) = f_0
  \frac{  v_F^2 \bar{k}^2 + \bar{\omega}^2}{ 
    v_c^2 \bar{k}^2 + \bar{\omega}^2}
  \; .
\end{equation}
Here $v_c = v_F \sqrt{ 1 + \tilde{f}_0}$ is the velocity of collective
charge excitations.  Equation (\ref{eq:sigmatruncres}) resembles the GW
approximation,\cite{Hedin65} but with the RPA interaction multiplied
by an additional singular vertex correction $ ( \Lambda_0 / |
{\bar{k}} |)^{ \eta}$.  The explicit evaluation of
Eq.~(\ref{eq:sigmatruncres}) is rather tedious and will not be further
discussed in this work.  The resulting spectral function $A ( k ,
\omega )$ agrees at $k = k_F$ with the bosonization result (even at
strong coupling), but has the wrong threshold singularities for $ |
\omega | \rightarrow v_c | ( k \pm k_F) | $.  So far we have not been
able to find a reasonably simple truncation of the exact flow
equations which completely produces the spectral line shape of $A ( k
, \omega )$, as predicted by bosonization or by our exact solution
presented in the previous section. Whether a self-consistent numerical
solution of the truncation discussed in Sec.~\ref{subsec:truncation}
[see Eqs.~(\ref{eq:flowSigmatrunc})-(\ref{eq:flowGammatrunc})] would
reproduce the correct spectral line shape or not remains an open
problem.  The numerical solution of these equations seems to be rather
difficult and is beyond the scope of this work.

\section{Summary and outlook}
\label{sec:summary}

In this work we have developed a new formulation of the functional RG
for interacting fermions, which is based on the explicit introduction
of collective bosonic degrees of freedom via a suitable
Hubbard-Stratonovich transformation.  A similar strategy has been used
previously in Refs.~\onlinecite{Correia01}, \onlinecite{Wetterich04},
and \onlinecite{Baier03}. However, on the technical level the
practical implementation of this method presented here differs
considerable from previous works.  We have payed particular attention
to asymptotic Ward identities, which play a crucial role if the
interaction is dominated by small momentum transfers.  In one
dimension, this is the key to obtain the exact solution of the
Tomonaga-Luttinger model entirely within the functional RG.  By using
the momentum transfer associated with the bosonic field as a cutoff
parameter, we have formulated the functional RG in such a way that the
RG flow does not violate the Ward identities. In fact, we have shown
that Ward identities emerge as the solution of the infinite hierarchy
of coupled RG flow equations for the one-line irreducible vertices
involving two external fermion legs and an arbitrary number of boson
legs. In principle this method can be iterated to obtain all
correlation functions of the TLM entirely within the framework of the
functional RG.

Here we have mainly laid the theoretical foundation of our approach
and developed an efficient method to keep track of all terms.  In
future work, we are planning to apply our technique to other
physically interesting problems.  Let us mention some problems where
it might be advantageous to use our approach:

(a) {\it{Strong coupling fixed points.}}  One of the big drawbacks of
the conventional (purely fermionic) functional RG used by many authors
\cite{Zanchi96,Halboth00,Honerkamp01,Honerkamp01b,Tsai01,Binz02,Meden02,Kampf03,Katanin04}
is that in practice the frequency dependence of the four-point vertex
$\Gamma^{(4)}$ has to be neglected, so that the wave-function
renormalization factor is $Z=1$.  Although the resulting runaway flow
of the vertices to strong coupling at a finite scale can be
interpreted in terms of corresponding instabilities, there is the
possibility that for small $Z$ the renormalized effective interaction
$Z^2\Gamma^{(4)}$ remains finite even though the vertex $\Gamma^{(4)}$
without wave-function renormalization seems to diverge.\cite{Ferraz03}
In our approach, the effective interaction acquires a
frequency dependence, even within the lowest-order approximation.  In
fact, if we ignore vertex corrections, the effective interaction is
simply given by the RPA.  Hence, strong coupling fixed points might be
accessible within our approach.  Recall that the rather simple
truncation of Sec.~\ref{subsec:truncrel} gave the exact anomalous
dimension of the TLM for arbitrary strength of the interaction.
Possibly, more elaborate truncations of the exact hierarchy of RG flow
equations (for example, the truncation based on retaining skeleton
elements of the two-point functions discussed in
Sec.~\ref{subsec:truncation}, see Fig.~\ref{fig:flowtrunc}) will give
accurate results for the spectral properties.

(b) {\it{Nonuniversal effects in one-dimensional metals.}}  If we do
not linearize the energy dispersion in one dimension, there should be
a finite momentum scale $k_c$ (depending on the interaction and the
band curvature) below which typical scaling behavior predicted by the
TLM emerges.  The calculation of $k_c$ as well as the associated
nonuniversal spectral line shape are difficult within
bosonization.\cite{Busche01} On the other hand, within the framework
of the functional RG the inclusion of irrelevant couplings is
certainly possible, so that with our method it might be possible to
shed some new light onto this old problem.  For an explicit
calculation of an entire crossover scaling function between the
critical regime and the short-wavelength regime of interacting bosons
in $D=3$, see Ref.~\onlinecite{Ledowski04}.  An analogous calculation
of the dynamic scaling functions for interacting fermions in one
dimension remains to be done.

(c) {\it{Itinerant ferromagnetism}.} Spontaneous ferromagnetism in
Fermi systems is driven by sufficiently strong interactions involving
small momentum transfers.  Assuming a given form of the ferromagnetic
susceptibility, Altshuler, Ioffe, and Millis \cite{Altshuler94}
concluded on the basis of an elaborate diagrammatic analysis that in
the vicinity of the paramagnetic-ferromagnetic quantum-critical point
in dimensions $D = 2$ a simple one-loop calculation already yields the
correct qualitative behavior of the electronic self-energy. If this is
true, then in this problem vertex corrections are irrelevant.
Unfortunately, due to the peculiar momentum and frequency dependence
of the ferromagnetic susceptibility $\chi ( \bar{{\bf{k}}} ,
\bar{\omega})$ at the quantum critical point, the assumption of asymptotic
velocity conservation [see Eq.~(\ref{eq:linearization}) in Appendix C]
leading to the Ward identity (\ref{eq:WI1})-(\ref{eq:WI2}) is not
justified.  Note, however, that in Ref.~\onlinecite{Altshuler94} the
form of the susceptibility is assumed to be given.  The feedback of
the collective ferromagnetic fluctuations on the non-Fermi-liquid form
of the electronic properties has not been discussed.  The fact that in
$D < 3$ the leading interaction corrections to the inverse
susceptibility $\chi^{-1} ( \bar{{\bf{k}}} , \bar{\omega})$ generate a
nonanalytic momentum dependence\cite{Belitz97,Chubukov03} suggests
that the problem should be reconsidered taking the interplay between
fermionic single-particle excitations and collective magnetic
fluctuations into account.  The formalism developed in this work might
be suitable to shed some new light also onto this problem.  The
electronic properties of a two-dimensional Fermi system in the
vicinity of a ferromagnetic instability have recently been studied in
Ref.~\onlinecite{Katanin04b}.  However, these authors focused on the
finite-temperature properties of the phase transition; they did not
attempt to calculate the fermionic single-particle Green's function in
the vicinity of the zero-temperature phase transition.  Note also that
for sufficiently strong interactions even one-dimensional fermions can
in principle have a ferromagnetic instability if the energy dispersion
is nonlinear.\cite{Bartosch03,Yang04}

(d) {\it{Quantum phase transitions and symmetry breaking.}}  Our
method unifies the traditional approach to quantum-phase transitions
pioneered by Hertz\cite{Hertz76} with the modern developments in the
field of fermionic functional RG, so that it might simplify the
theoretical description of quantum phase transitions in situations
where the fermions cannot be completely integrated out.  In order to
describe quantum phase transitions in interacting Fermi systems within
the framework of the traditional Ginzburg-Landau-Wilson approach, all
soft modes in the system should be explicitly
retained.\cite{Kirkpatrick96} In the purely fermionic functional RG
\cite{Zanchi96,Halboth00,Honerkamp01,Honerkamp01b,Salmhofer01,Kopietz01,Busche02,Ledowski03,Tsai01,Binz02,Meden02,Kampf03,Katanin04}
symmetry breaking manifests itself via the divergence of the relevant
order-parameter susceptibility; the symmetry broken phase is difficult
to describe within this approach. On the other hand, in our approach
the order parameter can be introduced explicitly as a bosonic field,
which acquires a vacuum expectation value in the symmetry broken
phase.  Previously, a similar approach has been developed in
Ref.~\onlinecite{Baier03} to study antiferromagnetism in the
two-dimensional Hubbard model.

In summary, we believe that the formulation of the exact functional RG
presented in this work will be quite useful in many different physical
contexts.

\section*{ACKNOWLEDGMENTS}

This work was completed while two of us (F. S. and P. K.) participated
at the {\it{Winter School on Renormalization Group Methods for
    Interacting Electrons}} at {\it{The International Center of
    Condensed Matter Physics (ICCMP)}} of the University of
Bras\'\i{}lia, Brazil.  This gave us the opportunity to discuss the
subtleties of the functional RG for Fermi systems with many colleagues
-- we thank all of them.  We also thank Alvaro Ferraz and the very
friendly staff of the ICCMP for their hospitality.  This work was
financially supported by the DFG, Grant No. KO 1442/5-3 (F. S. and
P. K.) and Grant No. BA 2263/1-1 (L. B.).


\begin{appendix}

\section{Tree expansion of connected Green's functions
  in terms of one-line irreducible vertices}
\label{sec:tree}

In this Appendix we show explicitly that the vertices $\Gamma^{(n)}_{
  \alpha_1 \ldots \alpha_n}$ defined in terms of the functional Taylor
expansion of $\Gamma [ \Phi ]$ in Eq.~ (\ref{eq:Gammaexpansion}) are
indeed one-line irreducible. This is usually \cite{Negele88} done
graphically by taking higher-order derivatives of the relation
(\ref{eq:quad_rel}) between the second functional derivatives of $ \LL
[ \Phi ]$ and $\G_c [ J ]$. With the help of our compact notation we
can even give the tree expansion of the connected Green's function in
terms of one-line irreducible vertices in closed form.  To do so, it
is advantageous to define the functional
\begin{equation}
  \mathbf{U} =\left[\ppt{\Gamma}{\Phi}{\Phi}-\left.\ppt{\Gamma}{\Phi}{\Phi}\right|_{\Phi=0}\right]^T
  =\left[\ppt{\Gamma}{\Phi}{\Phi}\right]^T-\mathbf{\Sigma}
  \; ,
  \label{eq:Udef}
\end{equation}
which is a matrix in superindex space.  With this definition, we have
\begin{equation}
  \ppt{\LL}{\Phi}{\Phi} = \mathbf{U}^T - [ \mathbf{G}^{-1} ]^T
  \label{eq:LUrelation}
  \; ,
\end{equation}
so that
\begin{eqnarray}
  \left[\ppt{\LL}{\Phi}{\Phi}\right]^{-1} & = & 
  -\mathbf{G}^T[\mathbf{1}-\mathbf{U}^T\mathbf{G}^T]^{-1}
  \nonumber
  \\
  & = & -\sum_{l=0}^{\infty}\mathbf{G}^T(\mathbf{U}^T\mathbf{G}^T)^l
  \; .
  \label{eq:Gc2}
\end{eqnarray}
From Eq.~(\ref{eq:quad_rel}) we then obtain
\begin{eqnarray}
  \ppt{\G_c}{J}{J} & = & \mathbf{Z}
  \left[ \ppt{\LL}{\Phi}{\Phi} \right]^{-1} =
  -  \mathbf{Z} \mathbf{G}^T[\mathbf{1}-\mathbf{U}^T\mathbf{G}^T]^{-1} 
  \nonumber
  \\
  & = & - \sum_{l=0}^{\infty} \mathbf{Z} \mathbf{G}^T(\mathbf{U}^T\mathbf{G}^T)^l
  \label{eq:GcJJexpansion}
  \; .
\end{eqnarray}
We now expand both sides of Eq.~(\ref{eq:GcJJexpansion}) in powers of
the sources $J$ and compare coefficients.  For the matrix on the
left-hand side we obtain, from Eq.~(\ref{eq:Gcexpansion}),
\begin{equation}
  \ppt{\G_c}{J}{J}
  =\sum_{n=0}^{\infty}\frac1{n!}\int_{\alpha_1}\dots
  \int_{\alpha_n} \left[\mathbf{G}^{(n+2)}_{c, \alpha_1 \dots \alpha_n}\right]^T J_{\alpha_1}
  \cdot\ldots\cdot J_{\alpha_n}\,,
  \label{eq:Gc2expansion}
\end{equation}
where the matrix $\mathbf{G}^{(n+2)}_{c, \alpha_1 \dots \alpha_n}$ is
defined by
\begin{equation}
  [ \mathbf{G}^{(n+2)}_{c, \alpha_1 \dots \alpha_n} ]_{\alpha  \alpha^{\prime}}
  = \G^{(n+2)}_{c, \alpha \alpha^{\prime}  \alpha_1 \dots \alpha_n }
  \; .
  \label{eq:Gcn2def}
\end{equation}
On the right-hand side we use Eqs.~(\ref{eq:Udef}) and
(\ref{eq:Gammaexpansion}) to write
\begin{equation}
  \mathbf{U} =\sum_{n=1}^{\infty}\frac1{n!}\int_{\alpha_1}\dots
  \int_{\alpha_n}  \mathbf{\Gamma}^{(n+2)}_{\alpha_1,\dots,\alpha_n}\Phi_{\alpha_1}
  \cdot\ldots\cdot\Phi_{\alpha_n}\,,
  \label{eq:Uexpansion}
\end{equation}
where
\begin{equation}
  [ \mathbf{\Gamma}^{(n+2)}_{\alpha_1,\dots,
    \alpha_n} ]_{\alpha \alpha^{\prime} }
  = \Gamma^{(n+2)}_{ \alpha \alpha^{\prime}  \alpha_1 \dots \alpha_n }
  \; .
  \label{eq:Gammamatrix}
\end{equation}
To compare terms with the same powers of the sources $J$ on both sides
of Eq.~(\ref{eq:GcJJexpansion}), we need to express the fields
$\Phi_\alpha$ on the right-hand side of Eq.~(\ref{eq:Uexpansion}) in
terms of the sources, using Eqs.~(\ref{eq:Gcexpansion}) and
(\ref{eq:def_phi}),
\begin{equation}
  \Phi_{\alpha}=\pp{\G_c}{J_{\alpha}}
  =
  \sum_{m=0}^{\infty}\frac1{m!}\int_{\beta_1}\dots
  \int_{\beta_m} {\G}^{(m+1)}_{c, \alpha \beta_1 \dots \beta_m} J_{\beta_1}
  \dots J_{\beta_m}
  \label{eq:PhiJexpansion}\,.
\end{equation}
Substituting Eqs.~(\ref{eq:Gc2expansion}), (\ref{eq:Uexpansion}), and
(\ref{eq:PhiJexpansion}) into Eq.~(\ref{eq:GcJJexpansion}) and comparing terms
with the same powers of the sources (after symmetrization), we obtain
a general relation between the connected and the one-line irreducible
correlation functions,
\begin{equation}
  \begin{array}{rcl}
    \mathbf{G}_{c,\beta_1,\dots,\beta_n}^{(n+2)}&=& -\sum\limits_{l=0}^{\infty}\sum\limits_{n_1,\dots,n_l=1}^{\infty}
    \frac{1}{n_1!\cdot\ldots\cdot n_l!}
    \\[0.4cm]
    & & \hspace{-12mm} \times
    \int_{\alpha_1^1}\dots\int_{\alpha_{n_1}^1}\dots\int_{\alpha_1^l}\dots\int_{\alpha_{n_l}^l}
    \\[0.4cm]
    &&   \hspace{-12mm} \times \sum\limits_{m_1^1,\dots,m_{n_1}^1=1}^{\infty}\dots\sum\limits_{m_1^l,\dots,m_{n_l}^l=1}^{\infty}
    \delta_{n,\sum_{i=1}^l\sum_{j=1}^{n_i}m^i_j}\\[0.8cm]
    &&  \hspace{-12mm} \times \left[\mathbf{Z}\mathbf{G}^T \mathbf{\Gamma}^{(n_1+2)\,T}_{\alpha_1^1,
        \dots,\alpha_{n_1}^1}\mathbf{G}^T\cdot\ldots\cdot\mathbf{G}^T
      \mathbf{\Gamma}^{(n_l+2)\,T}_{\alpha_1^l,\dots,\alpha_{n_l}^l}\mathbf{G}^T\right]^T
    \,\,\\
    &&  \hspace{-12mm} \times {\cal{S}}_{\beta_1,\dots,\beta_{m_1^1};\dots;\beta_{n-m^l_{n_l}+1},\dots,\beta_n}
    \Big\{    \G^{(m_1^1+1)}_{c,\alpha_1^1,\beta_1, \dots,\beta_{m_1^1}}   
    \\[0.4cm]
    && \hspace{-9mm}\cdot\ldots\cdot
    \G^{(m_{n_l}^l+1)}_{c,\alpha_{n_l}^l,\beta_{n-m^l_{n_l}+1},\dots,\beta_n}
    \Big\}
    \; .
  \end{array}
  \label{eq:GcGamma}
\end{equation}
On the right-hand side of this rather cumbersome expression, only
connected correlation functions with a degree smaller than on the
left-hand side appear. One can therefore recursively express all
connected correlation functions via their one-line irreducible
counterparts.  Only a finite number of terms contribute on the
right-hand side. The operator ${\cal{S}}$ symmetrizes the expression
in curly brackets with respect to indices on different correlation
functions, i.e., it generates all permutations of the indices with
appropriate signs, counting expressions only once that are generated
by permutations of indices on the same vertex. More precisely the
action of ${\cal{S}}$ is given by ($m=\sum_{i=1}^lm_i$)
\begin{eqnarray}
  {\cal{S}}_{\alpha_1,\dots,\alpha_{m_1};\dots;\alpha_{m-m_l+1},\dots,\alpha_{m}}
  \{ A_{\alpha_1,\dots,\alpha_m} \}
  & = & 
  \nonumber
  \\
  & & \hspace{-65mm} \frac{1}{\prod_i m_i!}\sum_P 
  \mathrm{sgn}_{\zeta}(P) \,
  A_{\alpha_{P(1)},\dots,\alpha_{P(m)}}\,,
  \label{eq:symmopdef}
\end{eqnarray}
where $P$ denotes a permutation of $\{1,\dots,m\}$ and
$\mathrm{sgn}_{\zeta}$ is the sign created by permuting field
variables according to the permutation $P$, i.e.,
\begin{equation}
  \Phi_{\alpha_1}\cdot\ldots\cdot\Phi_{\alpha_m} = 
  \mathrm{sgn}_{\zeta}(P) \,
  \Phi_{\alpha_{P(1)}}\cdot\ldots\cdot\Phi_{\alpha_{P(m)}}
  \; . 
\end{equation}
\begin{figure}
  \begin{center}
    \epsfig{file=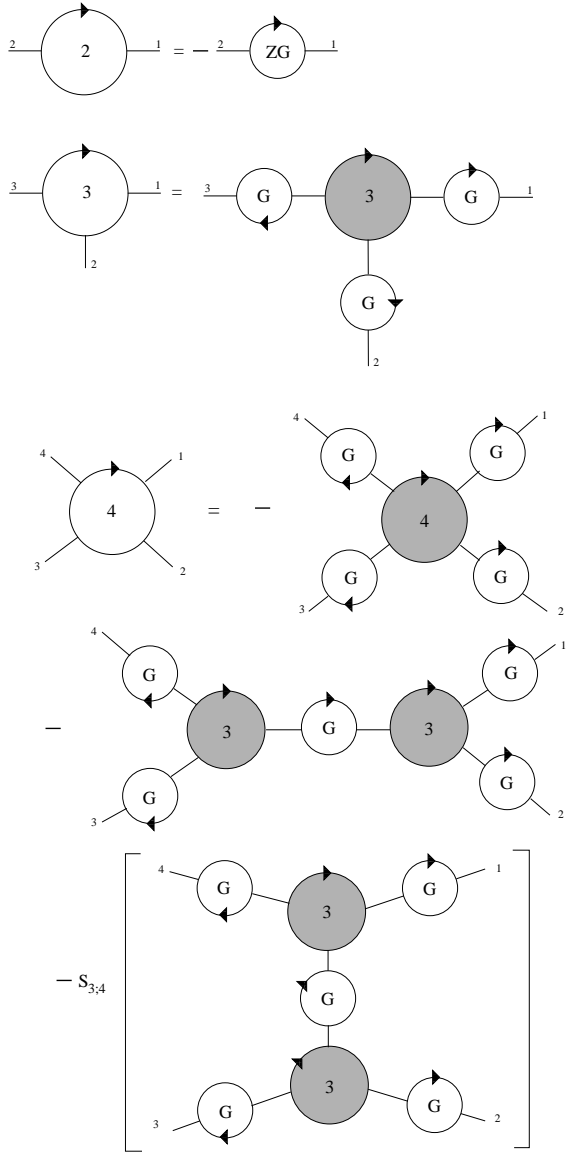,width=0.9\hsize}
  \end{center}
  \caption{Graphical representation of the relation between connected 
    Green's functions and one-line irreducible vertices up to the
    four-point functions.  The irreducible vertices are represented by
    shaded oriented circles with the appropriate number of legs; see
    Fig.~\ref{fig:generalvertex}.  The connected Green's functions are
    drawn as empty oriented circles with a number indicating the
    number of external legs.  }
  \label{fig:GcGamma}
\end{figure}
A diagrammatic representation of the first few terms of the
tree expansion generated by Eq.~(\ref{eq:GcGamma}) is given in
Fig.~\ref{fig:GcGamma}.  Let us give the corresponding analytic
expressions: If we set $n=0$ in Eq.~(\ref{eq:GcGamma}), then only the
term with $l=0$ contributes, and we obtain
\begin{equation}
  \mathbf{G}_c^{(2)} = - \mathbf{Z} \mathbf{G} = - \mathbf{G}^T
  \; ,
  \label{eq:Gtree2}
\end{equation}
which is Eq.~(\ref{eq:Gmatrix}) in matrix form.  For $n=1$ the single
term with $l=1$, $n_1 =1$, $m_1^1=1$ contributes on the right-hand
side of Eq.~(\ref{eq:GcGamma}).  Using $\mathbf{Z} \mathbf{G} =
\mathbf{G}^T$ the tree expansion of the connected Green's function with
three external legs can be written as
\begin{equation}
  \G^{(3)}_{c, \beta_1 \beta_2 \beta_3 }
  = \int_{\alpha_1} \int_{\alpha_2} \int_{\alpha_3}
  [\mathbf{G}]_{ \beta_1 \alpha_1}
  [\mathbf{G}]_{ \beta_2 \alpha_2}
  [\mathbf{G}]_{ \beta_3 \alpha_3}
  \Gamma^{(3)}_{ \alpha_1 \alpha_2 \alpha_3}
  \; .
\end{equation}
Finally, consider the connected Green's function with four external
legs, corresponding to $n=2$ in Eq.~(\ref{eq:GcGamma}).  In this case
the following three terms contribute:
\begin{eqnarray*}
  \nonumber
  \begin{array}{l|c|c|c}
    {\rm{term}} & l & n_i & m^i_j \\
    \colrule
    1.) &1  &n_1 =1 & m_1^1 =2 \\
    2.) & 1 & n_1 =2 & m_1^1 =m_2^1 = 1 \\
    3.) & 2 & n_1 = n_2 =1 & m_1^1 = m_1^2 =1  
  \end{array}
\end{eqnarray*}
The corresponding analytic expression is
\begin{eqnarray}
  \G^{(4)}_{c, \beta_1 \beta_2 \beta_3 \beta_4}
  & = &
  \nonumber
  \\
  & & \hspace{-20mm} - \int_{\alpha_1} \ldots \int_{\alpha_4}
  [\mathbf{G}]_{ \beta_1 \alpha_1}
  [\mathbf{G}]_{ \beta_2 \alpha_2}
  [\mathbf{G}]_{ \beta_3 \alpha_3}
  [\mathbf{G}]_{ \beta_4 \alpha_4}
  \Gamma^{(4)}_{ \alpha_1 \alpha_2 \alpha_3 \alpha_4}
  \nonumber
  \\
  & & \hspace{-20mm}
  - \int_{\alpha_1} \ldots \int_{\alpha_6}
  [\mathbf{G}]_{ \beta_1 \alpha_1}
  [\mathbf{G}]_{ \beta_2 \alpha_2}
  [\mathbf{G}]_{ \beta_3 \alpha_3}
  [\mathbf{G}]_{ \beta_4 \alpha_4}
  \nonumber 
  \\
  & & \times  \Gamma^{(3)}_{ \alpha_1 \alpha_2 \alpha_5}
  [\mathbf{G}]_{ \alpha_5 \alpha_6}
  \Gamma^{(3)}_{ \alpha_6 \alpha_3 \alpha_4}
  \nonumber
  \\
  & & \hspace{-20mm}
  - \int_{\alpha_1} \ldots \int_{\alpha_6}
  S_{\beta_3;\beta_4}  \Big\{  
  [\mathbf{G}]_{ \beta_1 \alpha_1}
  [\mathbf{G}]_{ \beta_2 \alpha_2}
  [\mathbf{G}]_{ \beta_3 \alpha_3}
  [\mathbf{G}]_{ \beta_4 \alpha_4}
  \nonumber 
  \\
  & & \times
  \Gamma^{(3)}_{ \alpha_1 \alpha_5 \alpha_4}
  [\mathbf{G}]_{ \alpha_5 \alpha_6}
  \Gamma^{(3)}_{ \alpha_6 \alpha_2 \alpha_3}
  \Big\}
  \; . 
  \label{eq:Gtree4}
\end{eqnarray}

\section{Dyson-Schwinger equations and skeleton diagrams}
\label{sec:skeleton}

In this appendix we show how the skeleton diagrams for the two-point
functions and the three-legged vertex in
Fig.~\ref{fig:skeletonsigmapi} can be formally derived from the
Dyson-Schwinger equations of motion.  Although the skeleton graphs are
usually written down directly from topological considerations of the
structure of diagrammatic perturbation theory,\cite{Nozieres64} it is
instructive to see how the skeleton expansion of the irreducible
vertices can be derived formally within our functional integral
approach.

The invariance of the generating functional $\G [ J] $ of the Green's
functions defined in Eq.~(\ref{eq:Ggen}) with respect to infinitesimal
shifts in the integration variables $\Phi_{\alpha}$ implies the
Dyson-Schwinger equations of motion \cite{ZinnJustin98}
\begin{equation}
  \left( \zeta_{\alpha} J_{\alpha} - 
    \frac{ \delta S}{\delta \Phi_{\alpha}} \left[ \frac{ \delta}{\delta J_{\alpha} } \right] 
  \right) \G [ J_{\alpha} ] = 0
  \; .
  \label{eq:DysonSchwinger}
\end{equation}
For our coupled Fermi-Bose system with Euclidean action $S [
\bar{\psi} , \psi , \varphi ]$ given by Eqs.~(\ref{eq:S0psidef},
\ref{eq:Spsiphidef}, \ref{eq:S0phi}, \ref{eq:Spsiphi}) involving three
types of fields, Eq.~(\ref{eq:DysonSchwinger}) is actually equivalent
with the following three equations:
\begin{eqnarray}
  & & 
  \Biggl(  J_{ - \bar{K} \sigma} - 
  \sum_{\sigma^{\prime}} [ f_{ \bar{\bf{k}} }^{-1} ]^{\sigma^{\prime} \sigma}
  \frac{ \delta}{ \delta J_{\bar{K} \sigma^{\prime}} }
  \Biggr)  \G
  \nonumber
  \\
  & & \hspace{10mm}
  - i \zeta \int_K \frac{ \delta^{(2)}  \G }{ 
    \delta j_{ K + \bar{K} \sigma} \delta \bar{\jmath}_{ K \sigma} } =0
  \; ,
  \label{eq:DysonSchwinger1}
  \\
  & &
  \Biggl( \zeta \bar{\jmath}_{ K \sigma} 
  + [ i \omega - \xi_{ {\bf{k}} \sigma } ]
  \frac{ \delta}{\delta j_{ K \sigma}}
  \Biggr) \G 
  \nonumber
  \\
  & & \hspace{10mm}
  - i \int_{\bar{K}} \frac{ \delta^{(2)}  \G }{ 
    \delta j_{ K + \bar{K} \sigma} \delta J_{ -\bar{K} \sigma} } =0
  \; ,
  \label{eq:DysonSchwinger2}
  \\
  & &
  \Biggl(  j_{ K \sigma} 
  + [ i \omega - \xi_{ {\bf{k}} \sigma } ]
  \frac{ \delta}{\delta \bar{\jmath}_{ K \sigma}}
  \Biggr) \G
  \nonumber
  \\
  & & \hspace{10mm}
  - i \int_{\bar{K}} \frac{ \delta^{(2)} \G }{ 
    \delta \bar{\jmath}_{ K - \bar{K} \sigma} \delta J_{ -\bar{K} \sigma} } =0
  \; .
  \label{eq:DysonSchwinger3}
\end{eqnarray}
Expressing these equations in terms of the generating functionals
$\G_c [ \bar{\jmath} , j , J]$ of the connected Green's functions and
the corresponding generating functional $\Gamma [ \bar{\psi} , \psi ,
\varphi ]$ of the irreducible vertices defined in Eq.~(\ref
{eq:Gammadef}), we obtain the Dyson-Schwinger equations of motion in
the following form:
\begin{eqnarray}
  & & 
  \frac{ \delta \Gamma}{\delta \varphi_{ \bar{K} \sigma} }
  - i \int_K \left[
    \bar{\psi}_{ K + \bar{K}, \sigma} \psi_{ K \sigma} +
    \frac{ \delta^{(2)} \G_c}{ \delta \bar{\jmath}_{ K \sigma} \delta j_{ K + \bar{K} , \sigma} }
  \right]  = 0 \; ,
  \nonumber
  \\
  \label{eq:DysonSchwinger4}
  \\
  & &
  \frac{ \delta \Gamma}{\delta \psi_{ {K} \sigma} }
  - i \int_{\bar{K}} \left[ \zeta
    \bar{\psi}_{ K + \bar{K}, \sigma} \varphi_{ \bar{K} \sigma} +
    \frac{ \delta^{(2)} \G_c}{  \delta j_{ K + \bar{K} , \sigma}  \delta J_{ - \bar{K} \sigma}}
  \right]  = 0 \; ,
  \nonumber
  \\
  \label{eq:DysonSchwinger5}
  \\
  & &
  \frac{ \delta \Gamma}{\delta \bar{\psi}_{ {K} \sigma} }
  - i \int_{\bar{K}} \left[ 
    {\psi}_{ K - \bar{K}, \sigma} \varphi_{ \bar{K} \sigma} +
    \frac{ \delta^{(2)} \G_c}{  \delta \bar{\jmath}_{ K - \bar{K} , \sigma}  \delta J_{ - \bar{K} \sigma}}
  \right]  = 0 \; .
  \nonumber
  \\
  \label{eq:DysonSchwinger6}
\end{eqnarray}
The second functional derivatives of $\G_c$ can be expressed in terms
of the irreducible vertices using Eq.~(\ref{eq:GcJJexpansion}).
Taking derivatives of
Eqs.~(\ref{eq:DysonSchwinger4}--\ref{eq:DysonSchwinger6}) with respect
to the fields and then setting the fields equal to zero we obtain the
desired skeleton expansions of the irreducible vertices.  Let us start
with the skeleton diagram for the self-energy shown in
Fig.~\ref{fig:skeletonsigmapi}(a). To derive this,we simply
differentiate Eq.~(\ref{eq:DysonSchwinger6}) with respect to $\psi_{
  K^{\prime} \sigma}$.  Using
\begin{equation}
  \left. \frac{ \delta^{(2)} \Gamma}{ \delta \psi_{ K^{\prime} \sigma} \delta 
      \bar{\psi}_{ K \sigma} }
  \right|_{ \rm{fields} = 0} = \delta_{ K , K^{\prime}} \Sigma_{\sigma} (K)
  \label{eq:Gamma2Sigma}
  \; ,
\end{equation}
we obtain
\begin{equation}
  \delta_{ K , K^{\prime}} \Sigma_{\sigma} (K) = i
  \int_{ \bar{K}} 
  \left. \frac{ \delta^{(3)} \G_c}{  \delta \psi_{ K^{\prime} \sigma}
      \delta \bar{\jmath}_{ K - \bar{K} , \sigma}  \delta J_{ - \bar{K} \sigma}}
  \right|_{ \rm{fields} = 0}
  \; .
\end{equation}
From the $l=1$ term in the expansion~(\ref{eq:GcJJexpansion}) it is easy to show that
\begin{eqnarray}
  \left. \frac{ \delta^{(3)} \G_c}{  \delta \psi_{ K^{\prime} \sigma}
      \delta \bar{\jmath}_{ K - \bar{K} , \sigma}  \delta J_{ - \bar{K} \sigma}}
  \right|_{ \rm{fields} = 0} =  \delta_{ K , K^{\prime}} F_{ \sigma \sigma}  ( \bar{K} ) 
  &&
  \nonumber
  \\
  \times G_{\sigma}  ( K+\bar{K} )
  \Gamma^{(2,1)} ( K + \bar{K} \sigma ; K \sigma ; \bar{K} \sigma )
  \; ,&&
\end{eqnarray} 
so that
\begin{eqnarray}
  \Sigma_{\sigma} (K) &=& i \int_{ \bar{K}}
  F_{ \sigma \sigma}  ( \bar{K} ) G_{\sigma}  ( K + \bar{K} )
  \nonumber\\
  &&~~~~\times\Gamma^{(2,1)}  ( K + \bar{K} \sigma ; K \sigma ; \bar{K} \sigma )
  \; ,
\end{eqnarray}
which is the analytic expression for the skeleton graph shown in
Fig.~\ref{fig:skeletonsigmapi}(a).  Similarly, we obtain the skeleton
expansion of the irreducible polarization by differentiating
Eq.~(\ref{eq:DysonSchwinger4}) with respect to $\varphi_{ - \bar{K}
  \sigma}$,
\begin{eqnarray}
  \Pi_{\sigma} ( \bar{K} ) & = &
  i  
  \int_{ {K}} 
  \left. \frac{ \delta^{(3)} \G_c}{  \delta \varphi_{ -\bar{K} \sigma}
      \delta \bar{\jmath}_{ K  , \sigma}  \delta j_{ K + \bar{K} \sigma}}
  \right|_{ \rm{fields} = 0}
  \nonumber
  \\
  &  & \hspace{-16mm} = - i \zeta \int_K 
  G_{\sigma} ( K ) G_{\sigma} ( K + \bar{K} )
  \Gamma^{(2,1)} ( K + \bar{K} \sigma ; K \sigma ; \bar{K} \sigma )
  \; ,
  \nonumber
  \\
  & &
\end{eqnarray} 
which is shown diagrammatically in Fig.~\ref{fig:skeletonsigmapi} (b).
Finally, applying the operator $\frac{\delta^{(2)}}{ \delta
  \bar{\psi}_{ K + \bar{K} \sigma} \delta \psi_{ K \sigma}}$ to
Eq.~(\ref{eq:DysonSchwinger4}) and subsequently setting the fields
equal to zero we obtain the skeleton expansion of the three-legged
vertex shown in Fig.~\ref{fig:skeletonsigmapi}(c),
\begin{eqnarray}
  \Gamma^{(2,1)} ( K + \bar{K} \sigma ; K \sigma ; \bar{K} \sigma )
  & = & i  
  \nonumber
  \\
  &  & \hspace{-45mm} - i  \zeta \int_{ K^{\prime}} 
  G_{\sigma} ( K^{\prime} ) G_{\sigma} ( K^{\prime} + \bar{K} )
  \nonumber
  \\
  & & \hspace{-37mm} \times
  \Gamma^{(4,0)} ( K + \bar{K} \sigma , K^{\prime}  \sigma ;
  K^{\prime} + \bar{K}  \sigma , K \sigma )
  \; .
  \label{eq:skeletongamma}
\end{eqnarray}
Skeleton expansions for higher-order vertices can be obtained
analogously from the appropriate functional derivatives of
Eqs.~(\ref{eq:DysonSchwinger4})--(\ref{eq:DysonSchwinger6}).

\section{Ward identities}
\label{sec:ward}

In this appendix we give a self-contained derivation of the Ward
identities in Eqs.~(\ref{eq:WI1})--(\ref{eq:WI2}) within the framework
of our functional integral approach.  Although the Ward identity
(\ref{eq:WI1}) for the three-legged vertex is well
known,\cite{Dzyaloshinskii74,Bohr81,Metzner98,Kopietz97} it seems that
the higher-order Ward identities given in Eqs.~(\ref{eq:WIm}) and
(\ref{eq:WI2}) cannot be found anywhere in the literature.  Since we
are interested in deriving infinitely many Ward identities involving
the vertices $\Gamma^{(2,m)}$ with two fermion legs and an arbitrary
number $m$ of boson legs, it is convenient to derive first a ``master
Ward identity'' for the generating functional for the irreducible
vertices, from which we can obtain all desired Ward identities for the
vertices by taking appropriate functional derivatives.

Consider the generating functional of the Green's function of our mixed
Fermi-Bose theory defined in Eq.~(\ref{eq:Ggen}), which in explicit
notation is given by
\begin{equation}
  \G[\bar{\jmath}, j , J]  = \frac{1}{\Z_0}  
  \int  D [ \bar{\psi} , \psi, \varphi ]
  e^{-S [ \bar{\psi} , \psi , \varphi ] +(\bar{\jmath} ,\psi) + ( \bar{\psi} , j )
    + ( J^*, \varphi )}
  \; .
  \label{eq:Ggenexplicit}
\end{equation}
If we rewrite the parts of the action involving the fermionic fields
$\bar{\psi}$ and $\psi$ in real space and imaginary time, the
Euclidean action reads [we use again the notation $X =
(\tau,{\bf{r}})$ introduced in Sec.~\ref{subsec:exactTLM}] as
\begin{eqnarray}
  S[ \bar{\psi} , \psi , \varphi ]
  &=& S_0[\bar{\psi},\psi] + S_0[\varphi] + S_1[\bar{\psi},\psi,\varphi]
  \nonumber\\
  S_0[\bar{\psi},\psi]
  &=& \sum_{\sigma} \int_X \bar{\psi}_{\sigma}(X)\partial_{\tau}\psi_{\sigma}(X)
  \nonumber\\
  &+&\sum_{\sigma}\int\!d\tau\int\!d^Dr\int\!d^Dr'\,\bar{\psi}_{\sigma}(\tau,\mathbf{r})
  \nonumber\\\
  &&~~~~~~~~~~\times
  \xi_{\sigma}(\mathbf{r}-\mathbf{r}')
  \psi_{\sigma}(\tau,\mathbf{r}')
  \;,\\
  S_1[\bar{\psi},\psi,\varphi]
  &=& i\sum_{\sigma}\int_X \bar{\psi}_{\sigma}(X)\psi_{\sigma}(X)\varphi_{\sigma}(X)\;,
  \label{eq:Slocal}
\end{eqnarray} 
where we have defined the Fourier transform of the dispersion
\begin{equation}
  \xi_{\sigma}(\mathbf{r})=\int\frac{d^Dk}{(2\pi)^D}\,\xi_{\mathbf{k}\sigma}\,e^{i\mathbf{k}\cdot\mathbf{r}}
  \;.
\end{equation}
Suppose now that we perform a local gauge transformation on the fermion
fields, defining new fields $\psi^{\prime}$ and $\bar{\psi}^{\prime}$
via
\begin{equation}
  \psi_{\sigma} ( X ) = e^{ i \alpha_{\sigma} ( X ) }
  \psi^{\prime}_{\sigma} ( X )
  \; \; , \; \;
  \bar{\psi}_{\sigma} ( X ) = e^{ - i \alpha_{\sigma} ( X ) }
  \bar{\psi}^{\prime}_{\sigma} ( X )
  \; ,
  \label{eq:gaugepsi}
\end{equation}
where $\alpha_{\sigma} ( X )$ is an arbitrary real function.  It is easy
to show that, to linear order in $\alpha_{\sigma} ( X )$, the action
(\ref{eq:Slocal}) transforms as follows:
\begin{eqnarray}
  S [ e^{ - i \alpha} \bar{\psi}^{\prime} , e^{i \alpha} \psi^{\prime} , 
  \varphi ] &=& 
  S [ \bar{\psi}^{\prime} , \psi^{\prime} , \varphi]
  \nonumber\\
  &&\hspace{-4cm}+\;i\sum_{\sigma}\int_X \bar{\psi}^{\prime}_{\sigma}(X)
  [\partial_{\tau}\alpha_{\sigma}(X)]\psi^{\prime}_{\sigma}(X)
  \nonumber\\
  &&
  \hspace{-4cm}-\;i\sum_{\sigma}\int\!d\tau\int\!d^Dr\int\!d^Dr'\,
  \bar{\psi}^{\prime}_{\sigma}(\tau,\mathbf{r})
[\alpha_{\sigma}(\tau,\mathbf{r})-\alpha_{\sigma}(\tau,\mathbf{r}')]
  \nonumber\\
  &&
  \times
  \xi_{\sigma}(\mathbf{r}-\mathbf{r}')\psi^{\prime}_{\sigma}(\tau,\mathbf{r}')
  \; .
  \label{eq:Sinvariance}
\end{eqnarray}
Using this relation, we see that the invariance of the generating
functional in Eq.~(\ref{eq:Ggenexplicit}) with respect to the change
of integration variables defined by Eq.~(\ref{eq:gaugepsi}) implies, to
linear order in $\alpha_{\sigma} ( X )$,
\begin{eqnarray}
  0 & = & 
  \frac{1}{\Z_0}  
  \int  D [ \bar{\psi} , \psi, \varphi ]
  e^{-S [ \bar{\psi} , \psi , \varphi ] +(\bar{\jmath} ,\psi) + ( \bar{\psi} , j ) n+ ( J , \varphi )}
  \nonumber
  \\
  & \times & \Biggl\{
  -\;\sum_{\sigma}\int_X \bar{\psi}_{\sigma}(X)
  [\partial_{\tau}\alpha_{\sigma}(X)]\psi_{\sigma}(X)
  \nonumber\\
  &&
  ~~+\;\sum_{\sigma}\int\!d\tau\int\!d^Dr\int\!d^Dr'\,
  \bar{\psi}_{\sigma}(\tau,\mathbf{r})
    \nonumber\\
  &&
  ~~~~~~~\times
  [\alpha_{\sigma}(\tau,\mathbf{r})-\alpha_{\sigma}(\tau,\mathbf{r}')]
  \xi_{\sigma}(\mathbf{r}-\mathbf{r}')\psi_{\sigma}(\tau,\mathbf{r}')
  \nonumber\\
  &&
  ~~+\;(\bar{\jmath},\alpha\psi)-(\bar{\psi}\alpha,j)
  \;\;
  \Biggr\} \; .
  \label{eq:invarianceG}
\end{eqnarray}
Taking the functional derivative of this equation with respect to
$\alpha_{\sigma}(X)$, this implies in Fourier space,
\begin{eqnarray}
  0 & = & 
  \int_K \Biggl\{  
  \left[ i \bar{\omega} - \xi_{\mathbf{k}+\bar{\mathbf{k}},\sigma} + \xi_{\mathbf{k}\sigma}
  \right]
  \frac{ \delta^{(2)} \G  }{ \delta \bar{\jmath}_{K \sigma} \delta j_{ K + \bar{K} \sigma} }
  \nonumber
  \\
  &    &  + 
  \bar{\jmath}_{ K + \bar{K} \sigma} 
  \frac{ \delta \G }{ \delta \bar{\jmath}_{ K \sigma} }
  - 
  j_{ K \sigma} \frac{ \delta \G }{ \delta j_{ K + \bar{K} \sigma}} 
  \Biggr\}
  \label{eq:WIG2}
  \; .
\end{eqnarray}
Expressing this equation in terms of the generating functional $\G_c =
\ln \G $ of the connected Green's functions and the generating
functional and $\Gamma [ \bar{\psi} , \psi , \varphi ]$ of the
irreducible vertices as defined in Eq.~(\ref{eq:Gammadef}), we obtain
\begin{eqnarray}
  0 & = & 
  \int_K \Biggl\{  
  \left[ i \bar{\omega} - 
    \xi_{\mathbf{k}+\bar{\mathbf{k}},\sigma} + \xi_{\mathbf{k}\sigma}
  \right]
  \frac{ \delta^{(2)} \G_c  }{ 
    \delta \bar{\jmath}_{K \sigma}  \delta j_{ K + \bar{K} \sigma} }
  \nonumber
  \\
  &    &  + \psi_{ K \sigma} 
  \frac{ \delta \Gamma }{ \delta {\psi}_{ K + \bar{K} \sigma} }
  - \bar{\psi}_{ K + \bar{K} \sigma} 
  \frac{ \delta \Gamma }{ \delta \bar{\psi}_{ K  \sigma} }
  \Biggr\}
  \label{eq:masterWI}
  \; .
\end{eqnarray}
Alternatively, using the Dyson-Schwinger
equation~(\ref{eq:DysonSchwinger4}), we may rewrite this as
\begin{eqnarray}
  0 & = & 
  i \bar{\omega} \left[  
    \frac{ \delta \Gamma}{ \delta \varphi_{ \bar{K} \sigma} } 
    - i  \int_K \bar{\psi}_{ K + \bar{K} \sigma} \psi_{K \sigma} \right]
  \nonumber
  \\
  &-  & i
  \int_K  
  (\xi_{\mathbf{k}+\bar{\mathbf{k}},\sigma}-\xi_{\mathbf{k}\sigma})
  \frac{  \delta^{(2)} \G_c  }{ 
    \delta \bar{\jmath}_{K \sigma}  \delta j_{ K + \bar{K} \sigma} }
  \nonumber
  \\
  &  +  &  i \int_K \Biggl[ \psi_{ K \sigma} 
  \frac{ \delta \Gamma }{ \delta {\psi}_{ K + \bar{K} \sigma} }
  - \bar{\psi}_{ K + \bar{K} \sigma} 
  \frac{ \delta \Gamma }{ \delta \bar{\psi}_{ K  \sigma} }
  \Biggr]
  \label{eq:masterWI2}
  \; .
\end{eqnarray}
Equations.~(\ref{eq:masterWI}) and (\ref{eq:masterWI2}) are our ``master
Ward identities'' from which we can now obtain Ward identities for the
vertices by differentiation.  For example, taking the derivative
$\frac{\delta}{\delta \varphi_{ - \bar{K} \sigma}}$ of
Eq.~(\ref{eq:masterWI2}) we obtain
\begin{equation}
  i \bar{\omega} \Pi_{\sigma} ( \bar{K} ) - 
  \Pi^c_{\sigma} ( \bar{K} )
   = 0
  \label{eq:continuity}
  \; ,
\end{equation}
where we have defined
\begin{eqnarray}
  \Pi^c_{\sigma} ( \bar{K} ) & = &
  - i \zeta \int_K
  (\xi_{\mathbf{k}+\bar{\mathbf{k}},\sigma}-\xi_{\mathbf{k}\sigma})
  G_{\sigma} ( K ) G_{\sigma} ( K + \bar{K} )
  \nonumber
  \\
  & & \times
  \Gamma^{(2,1)} ( K + \bar{K} \sigma ; K \sigma ; \bar{K} \sigma )
  \; .
\end{eqnarray} 
Equation~(\ref{eq:continuity}) is a relation between response functions,
which follows more directly from the equation of continuity.  If we
are interested in vertices involving at least one fermionic momentum
and if the momentum transferred by the interaction is small, our master
Ward identities can be further simplified.  Then all fermionic momenta
lie close to a given point ${\bf{k}}_{ F , \sigma}$ on the Fermi
surface so that Eqs.~(\ref{eq:masterWI}) and (\ref{eq:masterWI2})
become simpler if we assume asymptotic velocity conservation.  This
means that we replace under the integral sign
\begin{equation}
  \xi_{\mathbf{k}+\bar{\mathbf{k}},\sigma} - \xi_{\mathbf{k}\sigma}
  \rightarrow {\bf{v}}_{F, \sigma} \cdot \bar{\mathbf{k}}
  \; .
  \label{eq:linearization}
\end{equation}
This approximation amounts to the linearization of the energy
dispersion relative to the point ${\bf{k}}_{ F , \sigma}$ on the Fermi
surface. Using again Eq.~(\ref{eq:DysonSchwinger4}), our master Ward
identity becomes
\begin{eqnarray}
  0 & = & 
  (  i \bar{\omega} - {\bf{v}}_{F, \sigma}  \cdot \bar{\bf{k}} ) \left[  
    \frac{ \delta \Gamma}{ \delta \varphi_{ \bar{K} \sigma} } 
    - i  \int_K \bar{\psi}_{ K + \bar{K} \sigma} \psi_{K \sigma} \right]
  \nonumber
  \\
  &  +  &  i \int_K \Biggl[ \psi_{ K \sigma} 
  \frac{ \delta \Gamma }{ \delta {\psi}_{ K + \bar{K} \sigma} }
  - \bar{\psi}_{ K + \bar{K} \sigma} 
  \frac{ \delta \Gamma }{ \delta \bar{\psi}_{ K  \sigma} }
  \Biggr]
  \label{eq:masterWI3}
  \; .
\end{eqnarray}
Differentiating this simplified master Ward identity with respect to
the fields using the relation (\ref{eq:Gamma2Sigma}) as well as
\begin{eqnarray}
  \left .\frac{ \delta^{(3)} \Gamma}{ 
      \delta \varphi_{ \bar{K} \sigma } \delta \psi_{ K \sigma}
      \delta \bar{\psi}_{ K + \bar{K} \sigma } }
  \right|_{ \rm{fields} =0} &  &
  \nonumber \\
  & & \hspace{-40mm} =
  \Gamma^{(2,1)} ( K + \bar{K}\sigma  ; K \sigma ; \bar{K}\sigma  )
  \; ,
\end{eqnarray}
\begin{eqnarray}  
  \left .\frac{ \delta^{(4)} \Gamma}{ 
      \delta \varphi_{ \bar{K}_1 \sigma }
      \delta \varphi_{ \bar{K}_2 \sigma } \delta \psi_{ K \sigma}
      \delta \bar{\psi}_{ K + \bar{K}_1 + \bar{K}_2  \sigma } }
  \right|_{ \rm{fields} =0} &  &
  \nonumber \\
  & & \hspace{-60mm} =
  \Gamma^{(2,2)} ( K + \bar{K}_1 + \bar{K}_2\sigma  ; K \sigma; 
  \bar{K}_1 \sigma , \bar{K}_2 \sigma)
  \; ,
\end{eqnarray}
and so on, we obtain the Ward identities for the irreducible vertices
given in Eqs.~(\ref{eq:WI1}, \ref{eq:WIm}, \ref{eq:WI2}).

Of course, other Ward identities, e.g., the Ward identity for
$\Gamma^{(4,1)}$ discussed in Ref.~\onlinecite{Benfatto04}, can also
be obtained from Eq.~(\ref{eq:masterWI3}). Note that if the
approximation (\ref{eq:linearization}) is not made, the master Ward
identity (\ref{eq:masterWI3}) should be replaced by the more general
master Ward identity (\ref{eq:masterWI2}), so that the Ward identities
~(\ref{eq:WI1}, \ref{eq:WIm}, \ref{eq:WI2}) for the vertices acquire
correction terms. The effect of these correction terms on the Ward
identities for $\Gamma^{(2,1)}$ and $\Gamma^{(4,1)}$ has very recently
been studied in a mathematically rigorous way by Benfatto and
Mastropietro.\cite{Benfatto04}

\end{appendix}


\begin{thebibliography}{99}
\bibitem{footnotezeta}
The factor $\zeta $ arises from the antisymmetry of the Grassmann fields.
Although throughout this work it is understood that $\zeta =-1$,
all expressions  involving the factor $\zeta$
remain valid for bosonic fields $\psi$ and $\bar{\psi}$ if we set
$\zeta =+1$. We adopt here the notation of Ref.~\onlinecite{Negele88}.
\bibitem{footnotefield}
More precisely, we should distinguish the
quantum field $\Phi_{\alpha}$ from its expectation value, i.e.,  we should
write
$\langle \Phi_{\alpha} \rangle = \delta \G_c / \delta J_{\alpha}$.
For notational simplicity we redefine
$\langle \Phi_{\alpha} \rangle  \rightarrow \Phi_\alpha$.
Whenever we work with the Legendre effective action it is understood
that $\Phi_{\alpha}$ denotes the expectation value of the corresponding quantum field.
\bibitem{zfootnote} The factor of $\Omega_{\Lambda}^{-1}$ on the
  right-hand sides of Eqs.~(\ref{eq:Gammarescaledef}) and
  (\ref{eq:inhrescale1}) arises from the elimination of a fermionic
  frequency using the rescaled version of the energy-momentum
  conserving $\delta$ function
  $\delta_{K'_1+\ldots+K'_n,K_1+\ldots+K_n+\bar{K}_1+\ldots+\bar{K}_m}$.
    Such a procedure is only meaningful if $z_{\psi}\leq z_{\varphi}$,
    so that bosonic frequencies are not more relevant than fermionic
    ones. For $z_{\varphi}<z_{\psi}$ we should use the
    $\delta$ function to eliminate a bosonic frequency, which amounts
    to the replacement
    $\Omega_{\Lambda}^{-1}\rightarrow\bar{\Omega}_{\Lambda}^{-1}$ on
    the right-hand side of Eqs.~(\ref{eq:Gammarescaledef}) and
    (\ref{eq:inhrescale1}). This is the reason for the factor
    $z_{\text{min}}=\text{min}\{z_{\varphi},z_{\psi}\}$ in Eqs.~(\ref{eq:flowGammarescale1})
    and (\ref{eq:scaledim}).
\bibitem{footnoteVolker} V. Meden pointed out to us that this perfect
  agreement of Eq.~(\ref{eq:etares}) with bosonization might no longer
  occur in the more general TLM with $g_2\neq g_4$.
\end{thebibliography}
\end{document}